%% file: paper.tex
\documentclass[aip,jcp,preprint]{revtex4-1}
\usepackage[utf8]{inputenc}
\usepackage[rgb]{xcolor}             

\usepackage{amsmath}
\usepackage{braket}
\usepackage{amsfonts}
\usepackage{amssymb}
\usepackage{bm}
\usepackage{graphics}
\usepackage{natbib}
\usepackage{subcaption}
\usepackage{newfloat}
\usepackage{algorithm}
\usepackage{algpseudocode}

\usepackage{tikz}
\usetikzlibrary{arrows}
\usetikzlibrary{decorations.pathmorphing}
\usetikzlibrary{decorations.markings}
\usetikzlibrary{decorations.pathreplacing}

\newcommand{\field}{\varepsilon} 
\newcommand{\Tr}{\text{Tr}}     
\newcommand{\super}[1]{\mathcal{#1}} 
\newcommand{\commute}[2]{[#1, #2]} 

\renewcommand{\vec}{\boldsymbol}

\begin{document}

\title{An efficient spectral method for numerical time-dependent perturbation theory}
\date{\today}
\author{Cyrille Lavigne}
\author{Paul Brumer}
\affiliation{
  Chemical Physics Theory Group, Department of Chemistry,
  and Center for Quantum Information and Quantum Control, University of
  Toronto, Toronto, Ontario, M5S 3H6, Canada}

\begin{abstract}
  We develop the Fourier-Laplace Inversion of the Perturbation Theory
  (FLIPT), a novel numerically exact ``black box'' method to compute
  perturbative expansions of the density matrix with rigorous
  convergence conditions.  Specifically, the FLIPT method is extremely
  well-suited to simulate multiphoton pulsed laser experiments with
  complex pulse shapes.  The $n$-dimensional frequency integrals of the
  $n$-th order perturbative expansion are evaluated numerically using
  tensor products.  The $N$ points discretized integrals are computed in
  $O(N^2)$ operations, a significant improvement over the $O(N^n)$
  scaling of standard quadrature methods.
\end{abstract}
\maketitle

\section{Introduction}

At the microscopic level, the interaction between semiclassical light
and matter is well-described by perturbation
theory.\cite{shapiro_quantum_2012} Indeed, optical processes such as
two-photon absorption  and Raman scattering are often described and
classified in terms of discrete interactions with radiation, a picture
based on perturbation theory.\cite{cohen-tannoudji_atom-photon_1992}
Similarly, analytical results from perturbation theory are widely used
to analyse nonlinear
and ultrafast spectroscopic experiments,\cite{
  gallagher_faeder_two-dimensional_1999,
  zhuang_simulation_2006, quesada_effects_2014,
  bruhl_experimental_2018,reppert_classical_2018}
the light-induced control of quantum systems,\cite{
  pachon_mechanisms_2013,mukamel_coherent-control_2013,
  am-shallem_scaling_2014,lavigne_interfering_2017,
  lavigne_two-photon_2019}
and the dynamics of photoactivated natural
processes.\cite{tscherbul_excitation_2014,tscherbul_quantum_2015,
  brumer_shedding_2018,reppert_quantumness_2018}

Although perturbative analysis underlies much of the theory of optical
processes, it is not generally used in numerical simulations without
additional approximations.  For example, pulsed electric fields are
often approximated as an infinitely short $\delta(t)$
pulse,\cite{smallwood_analytical_2017,perlik_finite_2017} and nearly
monochromatic electric fields as infinitely long continuous wave (CW)
oscillations, i.e.  $\delta$ functions in frequency.  Both
approximations suffer from important drawbacks.  It is well-known that
the phase of an electric field affects the dynamics of matter
interacting with said field.\cite{shapiro_quantum_2012}  Such phase
effects range from the trivial reduction in two-photon absorption
probability when chirping ultrashort
pulses\cite{lavigne_two-photon_2019} to the non-trivial quantum
control of molecular
dynamics.\cite{
  brumer_laser_1992,mukamel_coherent-control_2013,
  am-shallem_scaling_2014,lavigne_interfering_2017,
  bruhl_experimental_2018,lavigne_two-photon_2019}
However, these effects are lost in the limit of infinitely short pulses.  Conversely,
simulations of interaction with CW radiation are of limited applicability, especially in the
presence of other time-dependent processes such as decoherence and
dissipation.\cite{spanner_communication_2010}   Furthermore, the
resulting time-independent equations do not converge without
renormalization\cite{faisal_theory_1987} or the inclusion of broadening
factors.\cite{mukamel_principles_1995,plenio_origin_2013}

In cases where an experimentally realistic electric field is required,
e.g., in the interpretation of control
experiments,\cite{lavigne_interfering_2017} the relevant Schrodinger
equation can be solved by means of a non-perturbative propagation of
the time-dependent Schrodinger equation, with the radiation included
in the Hamiltonian as a time-dependent
potential.\cite{katz_control_2010, arango_communication_2013,
abe_optimal_2005,bruhl_experimental_2018} A ``one-photon'' or
``two-photon'' result would be obtained in such a calculation by
choosing a small intensity for the electric field.

Time-dependent propagation methods, unlike perturbative approaches, do
not suffer from the lack of phase effects or convergence difficulties
described above.  However, direct propagation is non-perturbative,
which has important drawbacks when computing weak-field
processes. First, from a numerical analysis point of view, a high
accuracy is required to adequately simulate the small population
excited by the field.\cite{arango_communication_2013}  In addition,
short timesteps (sub-fs in the case of visible radiation) are required
to capture the fast oscillations of electric fields at optical
frequencies.  The rotating wave approximation (RWA) greatly speeds up
convergence by allowing for much longer timesteps.  However, the RWA
breaks down for non-resonant processes\cite{milonni_laser_2010} and
multiphoton absorption.\cite{faisal_theory_1987}

Second, non-perturbative results are difficult to interpret and to
compare with experimental results that are described in the framework
of perturbation theory, e.g., described by $n$-wave mixing or $n$
photon absorption processes.  For example, higher order processes can
appear in ``linear regime'' experiments and simulations
thereof. \cite{han_linear_2012,han_linear_2013,bruhl_experimental_2018,
lavigne_ultrafast_2019} In fact, non-perturbative microscopic
simulations (i.e. where only the wavefunction of a single molecule is
evolved) suffer in particular from uncontrolled contributions of
unwanted optical signals as they are not subject to the phase matching
conditions arising from the macroscopic Maxwell equations.  The final
computed result is then a sum of signals from processes that can only
be unmixed with great difficulty in
simulations.\cite{meyer_non-perturbative_2000}  In contrast, the
corresponding experimental signals are spatially separated.

In this article, a numerical algorithm is introduced to evaluate
arbitrary order perturbative expansions of the density matrix using
the Fourier-Laplace
inversion.\cite{
  dubner_numerical_1968,veillon_algorithm_1974,
  crump_numerical_1976,de_hoog_improved_1982,
  piessens_algorithm_1984}
The Laplace-transformed time-dependent perturbation theory is shown to
be convergent for experimentally relevant light-matter
processes. Perturbative contributions are obtained as iterated
frequency integrals.  A Fourier series discretization is used to
evaluate the perturbative integrals exponentially faster than standard
quadrature using a tensor product
technique.\cite{gerstner_numerical_1998}  The resultant
Fourier-Laplace Inversion of the Perturbation Theory (FLIPT) method
can be expressed succinctly using tensor notation and the associated
tensor product algebra is easily implemented using multidimensional
arrays.  The resultant, highly efficient implementation is made freely
available by the authors.\cite{lavigne_flipt.jl_2019}

The FLIPT algorithm introduced here can be used as a ``black box''
method to simulate light-matter interactions from
coherent,\cite{*[{Incoherent radiation, such as sunlight, is better
treated using other methods such as that described in }] [{}]
axelrod_efficient_2018} pulsed laser fields of the type used in
ultrafast spectroscopy.  The implementation is fully automatic and
contains no free parameters.  Since the computation is performed in
the frequency domain and not in the time domain, fast and slow
dynamical observables are equally resolved. This property is
particularly useful for chemical processes where coherent excitation
dynamics on the fs timescale lead to ps to ns reaction dynamics.  Both
are obtained in the FLIPT method in a numerically exact manner, with
uniform convergence and without the RWA.

Calculations below and in Ref.~\onlinecite{lavigne_two-photon_2019}
show that the FLIPT method can be applied to moderately large
multilevel systems (N $\approx$ 300-600) with timescales ranging from
the sub-fs dynamics of electronic coherences to the ps dynamics of
molecular vibrations. Furthermore, the method is not limited to
specific form for the exciting light and can equally treat resonant
and non-resonant multiphoton processes.  Extensions of the algorithm
to the perturbative analysis of wavefunction and non-Markovian
dynamics are discussed.  Convergence is rigorously established for
time-limited excitations of varying duration and is independent of the
underlying system
dynamics.\cite{dubner_numerical_1968,crump_numerical_1976}

\section{Theory}

The theory underlying the efficient numerical method alluded to above
is developed here.  The quantity to be computed is the perturbative
expansion of the density matrix under the action of time-dependent
potentials of the type relevant in ultrafast laser experiments.  The
usual time-dependent perturbative expansion of the density matrix is
re-derived to show that the Laplace transform formalism used here is
convergent and well-behaved.  Below, the perturbative expansion and
its computation through a Laplace inversion is demonstrated for the
density matrix obeying the Liouville-von Neumann equation.  Possible
extensions to wavefunction dynamics and open system dynamics are
briefly described.

The Hamiltonian of interest is given by,
\begin{align}
  H(t) = H_0 + \lambda E(t) V,
\end{align}
where $H_0$ is the zeroth-order Hamiltonian, $\lambda$ is a small
dimensionless scalar and $V$ is the coupling operator for the
perturbation that evolves under $E(t)$, a scalar function of
time.\footnote{The algorithm can easily be extended to the case where
the perturbation is composed of multiple components $\sum_\alpha
E_\alpha(t) V_\alpha(t)$ by computing and summing over all unique
combinations of $\alpha$.}  Significantly, such an Hamiltonian
describes the interaction of a molecule with a time-varying classical
electric field in the dipole approximation,\cite{shapiro_quantum_2012}
with
\begin{align}
  \lambda E(t) V = - \sum_\alpha \field_\alpha(\vec r, t) \mu_\alpha,
\end{align}
where $\vec r$ is the position of the molecule, $\field_\alpha(\vec r,
t)$ is the $\alpha=x,y,z$ component of the electric field and
$\mu_\alpha$ is the $\alpha$ component of the dipole transition
operator for the molecule.

The FLIPT method depends crucially on the convergence of the Laplace
transform applied to the perturbation, which is guaranteed in the
experimentally relevant case of a perturbation of finite duration, as
shown in this section.  Specifically, the perturbation $E(t)$ is taken
to be bounded and time-limited, i.e., $E(t) = 0$ for all $t$ less than
some ``turn-on time'' $t_\text{on}$ and larger than some ``turn-off
time'' $t_\text{off}$.  This condition guarantees convergence, both
analytically and numerically, as shown below.

\subsection{Perturbative expansion and Laplace inversion}
The Liouville-von Neumann equation of motion for the density
matrix in the superoperator formalism is given
by,\cite{lendi_superoperator_1977}
\begin{align}
  \frac{\mathrm d}{\mathrm d t} \rho(t) &= \super L_0 \rho(t) + \lambda E(t)\super V \rho(t)  \label{eq:liouville}\\
  \super L_0 \rho &= \frac{1}{i\hbar}\commute{H_0}{\rho}\label{eq:superL}\\
  \super V \rho &= \frac{1}{i\hbar}\commute{V}{\rho}\label{eq:superV},
\end{align}
where Liouvillian and coupling superoperators are so defined.  A
Fourier inversion of $E(t)$ yields the following,
\begin{align}
  \frac{\mathrm d}{\mathrm d t} \rho(t) &= \super L_0 \rho(t) + \frac{\lambda}{2\pi} \int_{-\infty}^\infty \mathrm d \omega' E(\omega') e^{i\omega' t}\super V \rho(t).
\end{align}
Without loss of generality, the origin $t=0$ is chosen before the ``turn-on time'' $t_\text{on}$.  A Laplace transform
yields,\cite{boas_mathematical_2005}
\begin{align}
  (z-\super L_0)\rho(z) &= \rho_0 + \frac{\lambda}{2\pi} \int_{-\infty}^\infty \mathrm d \omega' E(\omega') \super V \rho(z-i\omega').
\end{align}
where
\begin{align}
  \rho(z) = \int_{0}^\infty \mathrm d t e^{-z t} \rho(t),
\end{align}
and $\rho_0$ is the initial state $\rho(t=0)$.  Multiplying through by
the Green's function $\super G_0(z) = (z - \super L_0)^{-1}$, an
implicit integral equation for $\rho(z)$ is
obtained,\cite{lowdin_operators_1982}
\begin{align}
  \rho(z) &= \super G_0(z)\left(\rho_0 + \frac{\lambda}{2\pi} \int_{-\infty}^\infty \mathrm d \omega' E(\omega') \super V \rho(z-i\omega')\right).\label{eq:laplace-transformed-lvn} 
\end{align}
The perturbative expansion results from an iteration over $\rho(z)$:
\begin{align}
  \rho(z) &= \super G_0(z)\rho_0 + \frac{\lambda}{2\pi} \int_{-\infty}^\infty \mathrm d \omega' E(\omega') \super G_0(z)\super V \super G_0(z-i\omega') \rho_0\\
          &+  \left(\frac{\lambda}{2\pi}\right)^2 \iint_{-\infty}^\infty \mathrm d \omega' \mathrm d \omega'' E(\omega'') E(\omega') \super G_0(z)\super V \super G_0(z-i\omega') \super V \super G_0(z-i\omega' -i\omega'') \rho_0 + O(\lambda^3)\nonumber\\
  &= \rho_0(z) + \rho_1(z) + \rho_2(z) + \cdots
\end{align}
The Laplace-transformed $\rho_n(z)$ is the $n$-th order perturbative
contribution to $\rho(z)$.  The iteration procedure can be written
explicitly and succinctly as,
\begin{align}
  \rho_0(i\omega + \eta) &= \super G_0(i\omega + \eta) \rho_0 \label{eq:rec1}\\
  \rho_n(i\omega + \eta) &= \frac{\lambda}{2\pi} \int_{-\infty}^\infty \mathrm d \omega' E(\omega') \super G_0(i\omega+\eta)\super V \rho_{n-1}(i\omega-i\omega' +\eta). \label{eq:rec2}
\end{align}
Thus, $\rho_{n}(i\omega + \eta)$ is
obtained from a convolution of $E(\omega) \super V$ with
$\rho_{n-1}(i\omega - i\omega' + \eta)$ followed by an application of
$\super G_0(i\omega + \eta)$.

An important special case is where the system is initially in a steady
state, such that $\super L_0 \rho_0 = 0$.  Then, the first order
contribution becomes,
\begin{align}
  \rho_1(i\omega + \eta) &= \frac{\lambda}{2\pi} \int_{-\infty}^\infty \mathrm d \omega' E(\omega') \super G_0(i\omega+\eta)\super V \super G_0(i\omega - i\omega' + \eta) \rho_0 \\
                           &= \frac{\lambda}{2\pi}\super G_0(i\omega + \eta)\super V \rho_0 \int_{-\infty}^\infty \mathrm d \omega'  \frac{E(\omega')}{i\omega - i\omega' + \eta} .\label{eq:ss_0}
\end{align}
The integral over $\omega'$ is analytically solvable.  The one-sided
Fourier transform of an exponentially decaying function is given by,
\begin{align}
  \int_{-\infty}^{\infty} \mathrm d t e^{-i\omega t - \eta t} \Theta(t) =  \frac{1}{i\omega + \eta}.
\end{align}
where $\Theta(t)$ is the Heaviside step function.  The integral in
eq.~(\ref{eq:ss_0}) is thus given by,
\begin{align}
  \int_{-\infty}^\infty \mathrm d \omega'  \frac{E(\omega')}{i\omega - i\omega' + \eta} &= \int_{-\infty}^\infty \mathrm d \omega'E(\omega')\int_{-\infty}^\infty\mathrm d t e^{-i \omega t + i\omega' t - \eta t} \Theta(t) \\
  &= 2\pi \int_{-\infty}^\infty\mathrm d t e^{-i \omega t - \eta t}\Theta(t) E(t).\label{eq:integral}
\end{align}
Since the perturbation $E(t)$ is zero for $t<0$,
eq.~(\ref{eq:integral}) is the Fourier transform of the following
function,
\begin{align}
  E_\eta(t) = E(t) e^{-\eta t}.
\end{align}
Therefore, the first order term in eq.~(\ref{eq:rec2}) becomes,
\begin{align}
  \rho_1(i\omega + \eta) &= 2\pi E_\eta(\omega) \frac{\lambda}{2\pi}\super G_0(i\omega + \eta)\super V \rho_0
\end{align}
A convenient form for the $n$-th order term can be obtained by
introducing an additional frequency integration,
\begin{align}
 \int_{-\infty}^\infty \mathrm d \Omega \delta(\Omega - \omega),
\end{align}
and performing the change of variables $\Omega = \omega_1 + \omega_2
\cdots$, where $\omega_i$ is the frequency variable for the $i$-th
perturbation.  The
$n$-th perturbative contribution is then given by a $n$-dimensional
frequency integral, obtained from $n$ perturbations $E(\omega_1),
E(\omega_2) \cdots$ of the initial state $\rho_0$,
\begin{align}
  \rho_n(i\omega + \eta) &= 2\pi  \left(\frac{\lambda}{2\pi}\right)^{n}\int_{-\infty}^{\infty}\mathrm d \omega_n  \cdots\int_{-\infty}^{\infty}\mathrm d \omega_1 E(\omega_n) \cdots E(\omega_2) E_\eta(\omega_1 ) \delta\left(\omega - \sum^{n}_{i=1}\omega_i\right)\label{eq:rhon_ss}\\
                         &\times \super G_0(i\omega_n + \cdots +i\omega_1 + \eta) \super V \super G_0(i\omega_{n-1}+\cdots+i\omega_1 +  \eta) \cdots\nonumber\\
  & \times \super V\super G_0(i\omega_1  +  \eta)\super V \rho_0.\nonumber
\end{align}
The case of a system initially in a steady state is particularly
important, as that describes most spectroscopic and control
experiments.\cite{lavigne_ultrafast_2019}  It is also significantly
simpler as it yields time-translationally invariant dynamics, that is
dynamics that do not depend on the absolute value of the initial time
$t_\text{on}$.\footnote{
  The case where the initial state is not a steady state can be
  computed directly from eq.~(\ref{eq:rec1}) and (\ref{eq:rec2})
  above. This is numerically more expensive as it lead to an $n+1$
  dimensional integral for the $n$-th perturbative term instead of an
  $n$ dimensional integral.}

The time-dependent $n$-th order contribution $\rho_n(t)$ can be
obtained by the Laplace inversion integral,
\begin{align}
 \rho_n(t) = \frac{1}{2\pi i}\int^{\eta + i\infty}_{\eta - i\infty} \mathrm d z e^{z t}\rho_n(z) = \frac{e^{\eta t}}{2\pi}\int^{\infty}_{ - \infty} \mathrm d \omega e^{i\omega t }\rho_n(i\omega + \eta),\label{eq:inv-laplace}
\end{align}
where $z=i\omega + \eta$ and $\eta$ is a real number greater than the
real part of all the poles of $\rho_n(z)$.  Substituting
eq.~(\ref{eq:rhon_ss}) into eq.~(\ref{eq:inv-laplace}) yields the
following $n$ multidimensional inverse Fourier transform solution for
$\rho_n(t)$,
\begin{align}
  \rho_n(t) &= e^{\eta t}\left(\frac{\lambda}{2\pi}\right)^{n} \int_{-\infty}^{\infty}\mathrm d \omega_n  \cdots\int_{-\infty}^{\infty}\mathrm d \omega_1 \exp\left(i\sum_{i=1}^n \omega_i t\right)E(\omega_n) \cdots E(\omega_2) E_\eta(\omega_1)\label{eq:rhon_of_t} \\
                         &\times \super G_0(i\omega_n + \cdots + i\omega_1 + \eta) \super V \super G_0(i\omega_{n-1}+\cdots+i\omega_1 + \eta) \cdots \super V\super G_0(i\omega_1 + \eta)\super V\rho_0.\nonumber  
\end{align}
For a closed multilevel system, all eigenvalues of the Liouvillian
superoperator $\super L_0$ have a zero real part; thus all poles of
$\super G_0(z)$ lie on the real line.\cite{lowdin_operators_1982}  As
$E(\omega)$ is an entire function, this integral converges for any
positive value of $\eta$.  Numerically, as described below, a value of
$\eta$ that minimizes numerical error is used.

\subsection{Extensions to open system and wavefunction formalisms}
The perturbative analysis described above is readily extended to
general open system dynamics and to wavefunction calculations.  Below,
such extensions are briefly discussed.

Perturbative expansions of the wavefunction, which are significantly
more efficient to compute for large closed
systems,\cite{rose_numerical_2019} also admit a Laplace solution.  The
Schrodinger equation describing the evolution of the wavefunction
under the action of a time-dependent potential is given by,
\begin{align}
  \frac{\mathrm d}{\mathrm d t} \ket{\psi(t)} &= \frac{1}{i\hbar} H_0 \ket{\psi(t)} + \frac{1}{i\hbar} \lambda E(t) \mu \ket{\psi(t)}.\label{eq:schrodinger}
\end{align}
This equation is identical to eq.~(\ref{eq:liouville}) when the
following substitutions are performed,
\begin{align}
  \super L_0 &\rightarrow \frac{1}{i\hbar} H_0\\
  \super V &\rightarrow \frac{1}{i\hbar} \mu\\
  \rho_n(t) &\rightarrow \ket{\psi_n(t)}.
\end{align}
It is thus unsurprising that a near identical equation to
eq.~(\ref{eq:rhon_of_t}) is obtained upon performing the Laplace
transform and inversion,
\begin{align}
  \ket{\psi_n(t)} &= e^{\eta t}\left(\frac{\lambda}{2\pi}\right)^{n} \int_{-\infty}^{\infty}\mathrm d \omega_n  \cdots\int_{-\infty}^{\infty}\mathrm d \omega_1 \exp\left(i\sum_{i=1}^n \omega_i t\right)\\
  & \times E(\omega_n) \cdots E(\omega_2) E_\eta(\omega_1-\epsilon_0/\hbar)\nonumber \\
                         &\times  G_0(i\omega_n + \cdots + i\omega_1 + \eta)  V  G_0(i\omega_{n-1}+\cdots+i\omega_1 + \eta) \cdots  V G_0(i\omega_1 + \eta) V\ket{\psi_0},\nonumber  
\end{align}
where
\begin{align}
  G_0(i\omega + \eta) &= \frac{1}{i\omega +\eta - H_0/i\hbar}\\
  V &= \frac{1}{i\hbar} \mu,
\end{align}
and $\ket{\psi_0}$ is an eigenstate of $H_0$ with eigenenergy
$\epsilon_0$.  The term $E_\eta(\omega_1-\epsilon_0/\hbar)$ is
obtained from eq.~(\ref{eq:ss_0}) for the initial state with
$G_0(i\omega+\eta) \ket{\psi_0} = (i\omega + \eta
-\epsilon_0/i\hbar)^{-1} \ket{\psi_0}$.  While the wavefunction
approach can be computationally more efficient than the density matrix
approach for large closed systems, the perturbative expansion suffers
from well-known issues when it is used to compute expectation
values.\cite{mukamel_principles_1995}  For example, a second order
expansion of the wavefunction $\ket{\psi(t)} = \ket{\psi_0} +
\ket{\psi_1(t)} + \ket{\psi_2(t)}$ yields nine terms when taking the
expectation value $\braket{\psi(t)|O|\psi(t)}$, with perturbative
orders ranging from zero to four and with mixed time ordering.  In
contrast, the density matrix perturbation treat the state (the density
matrix) and the observables on an equal
footing.\cite{lavigne_two-photon_2019}

The FLIPT method can also be used to compute perturbative expansions
of open systems.  For example, open system dynamics of the Lindblad
type have been used by the authors in a study of two-photon control,
using the extension described here.\cite{lavigne_two-photon_2019}
Consider for instance the generalized master equation,
\begin{align}
  \frac{\mathrm d}{\mathrm d t} \rho(t) &= \int_{0}^{t} \mathrm d t' \super K(t-t') \rho(t') + \lambda E(t)\super V \rho(t),  \label{eq:nonmarkov}
\end{align}
where $\super K(t-t')$ is the memory kernel
superoperator.\cite{de_vega_dynamics_2017}  A Laplace transform yields
a similar expression to eq.~(\ref{eq:laplace-transformed-lvn}),
\begin{align}
  \rho(z)  = \left[z - \super K(z)\right]^{-1} \left(\rho_0  + \frac{\lambda}{2\pi}\int_{-\infty}^\infty \mathrm d \omega' E(\omega')  \super V \rho(z - i \omega')\right),
\end{align}
provided of course that the Laplace transform of the kernel exists,
\begin{align}
  \super K(z) = \int_0^{\infty}\mathrm d t e^{-z t}\super K(t).
\end{align}
Then, a perturbative expansion leads to a set of equations identical
to eqs.~(\ref{eq:rec1}) and (\ref{eq:rec2}) above other than having a
different Green's function,
\begin{align}
  \super G_0(i\omega + \eta) = \frac{1}{i\omega -  \super K(z) + \eta}.
\end{align}
For the case of a Markovian environment, the memory kernel is in fact
memory-less,
\begin{align}
  \super K(t-t') = \delta(t-t') (\super L_0  + \super R),
\end{align}
where $\super R$ is the relaxation tensor from e.g., the Redfield
equation.\cite{weiss_quantum_2012}  Then, an identical equation to the
closed system case above is obtained, with the exception that the
eigenvalues of $\super L_0$ are no longer strictly imaginary but
include negative real components.\cite{albert_symmetries_2014-2}
Evaluating $\super G_0(i\omega + \eta) \rho$ is more difficult but
those not affect the overall convergence; the additional broadening of
the spectra from decay and decoherence processes makes the integral of
eq.~(\ref{eq:rhon_ss}) better behaved than in the closed system case.
The case of non-Markovian dynamics would follow the same approach but
requires further consideration as to convergence and is beyond the
scope of this paper.

\section{The FLIPT method}
The FLIPT algorithm, introduced below, provides a highly efficient
method to numerically evaluate terms of the perturbative series.  The
numerical inversion of the Laplace transform is performed using a
well-known and well-understood Fourier series
approach.\cite{
  dubner_numerical_1968,veillon_algorithm_1974,
  crump_numerical_1976,de_hoog_improved_1982,
  piessens_algorithm_1984}
The Fourier inversion corresponds to using a finite difference grid in
the frequency domain, which is simple to implement and has well-known
error properties.  Importantly, this simple discretization scheme can
be used to exploit the iterative structure of the multidimensional
frequency domain integrals of eq.~(\ref{eq:rhon_ss}) and thereby
greatly reduce the number of integrand evaluations required.

Below, we describe a Fourier series discretization of the Laplace
inversion integral of eq.~(\ref{eq:rhon_of_t}).  The discretized
integral is expressed as a product of tensors.  In the second section
below, we show how this tensor form is exploited to obtain the FLIPT
algorithm. The numerical complexity and error properties of this
algorithm are then described.  Finally, we address the numerical
evaluation of $\super V \rho$ and $\super G_0(\omega) \rho$ and
related performance considerations.

\subsection{Tensor product representation}
The Laplace transform can be inverted using a Fourier series
decomposition in the time domain.\cite{dubner_numerical_1968}
Applying a finite difference discretization to
eq.~(\ref{eq:rhon_of_t}) gives the following approximation to $\rho_n(t)$,
\begin{align}
  \rho_n(t) &= \frac{e^{\eta t}}{2\pi} \sum_{k=-\infty}^{\infty} e^{i \Omega k t}\rho_n(\Omega k - i\eta).\label{eq:rhont-disc}
\end{align}
That is, the Laplace inversion has been approximated by its discrete
Fourier series over an interval of size $T$, with a corresponding
frequency $\Omega =
2\pi/T$.\cite{dubner_numerical_1968,crump_numerical_1976}  If $T$ is
longer than the length of the field $t_\text{off}-t_\text{on}$,
$E_\eta(t)$ can be exactly represented by its Fourier series transform
over the interval $T$. Without loss of generality (as described below)
the interval is taken here to be $[0,T]$.  Then, the Fourier series
for $E_\eta(t)$ is given by
\begin{align}
  E_\eta(t) = \sum_{k=-\infty}^{\infty} E_\eta[k] e^{ i \Omega  k t},\label{eq:fourier-series-t}
\end{align}
with Fourier coefficients given by,\cite{boas_mathematical_2005}
\begin{align}
  E_\eta[k] = \frac{1}{T}\int_0^T\mathrm d t e^{-i\Omega k t -\eta t}E(t). \label{eq:fourier-series}
\end{align}
The subscript $\eta$ will only be included for those term where $\eta
\neq 0$.  Since the Fourier series approximation to $E(t)$ is periodic,
it has the following Dirac comb as its Fourier transform:
\begin{align}
  E_\eta(\omega) = \int_{-\infty}^{\infty} \mathrm d t e^{-i\omega t} \sum_{k=-\infty}^{k=\infty} E_\eta[k] e^{i \Omega k t} = \sum_{k=-\infty}^{k=\infty} E_\eta[k] \delta(\omega - \Omega k).\label{eq:dirac-comb}
\end{align}
Substituting eq.~(\ref{eq:dirac-comb}) into eq.~(\ref{eq:rhon_ss}) yields a frequency-discretized expression,
\begin{align}                   
  \rho_n(t) &= \left(\frac{\lambda}{2\pi}\right)^{n} e^{\eta t}    \sum^\infty_{k_n=\infty}\cdots \sum^\infty_{k_1=\infty} \label{eq:flipt-main}\\
                   &\times\exp\left(i \Omega \sum_{i=1}^n k_i t\right) E[k_n] \cdots E[k_2] E_\eta[k_1]\nonumber\\
                         &\times \super G_0\left(i\Omega \sum^n_{i=1} k_i + \eta\right) \super V \super G_0\left(i\Omega \sum^{n-1}_{i=1}k_i+ \eta\right) \cdots\nonumber\\
  & \times \super V\super G_0(i\Omega k_1 + \eta)\super V \rho_0,\nonumber  
\end{align}
where $k_i$ is an integer index for the grid points of the discretized
integral over $\omega_i$.  This equation, the discretized
integral that yields the $n$-th order perturbative contribution to
$\rho(t)$, is at the core of the FLIPT algorithm.  Formulas are given
in the Appendix for the computation of spectral quantities.

As previously stated, the start of the interval over which the field
is on does not need to be explicitly included.  This property is a
consequence of the steady initial state of the system.  Consider the
value of $\rho_n(t)$ due to the translated perturbation $E'(t) = E(t -
t_0)$. The translated perturbation yields the following Fourier
series,
\begin{align}
  E'_\eta[k] &= \frac{1}{T}\int_{t_0}^{t_0 + T}\mathrm d t e^{-i\Omega k t - \eta t}E(t - t_0) \\
             &= e^{-i\Omega k t_0 - \eta t_0}\frac{1}{T}\int_0^T\mathrm d t e^{-i\Omega k t + \eta t}E(t) \\
             &= e^{-i\Omega k t_0 - \eta t_0} E_\eta[k].
\end{align} 
That is, the translation yields an additional factor of $e^{-i\Omega k
t_0 - \eta t_0}$ for each $E_\eta[k_i]$.  These factors generate a
corresponding translation of $\rho_n(t)$ to $\rho_n(t - t_0)$ by
acting on the exponential time-dependence of
eq.~(\ref{eq:rhont-disc}). Hence, $\rho_n(t-t_0)$ due to $E(t-t_0)$
equals $\rho_n(t)$ due to $E(t)$, as is expected from the steady-state
initial condition.

Importantly, this invariance to time-translations removes the need to
explicitly provide the interval over which the Fourier series of
$E(t)$ is computed --- this
information is entirely encoded in the function $E_\eta[k]$.  Thus, the only
parameters are $T$, the duration of the propagation of $\rho_n(t)$
after the field is on, and the convergence
parameter $\eta>0$ which, as shown below, can be expressed in terms of $T$.  

The specific structure of equation (\ref{eq:flipt-main}) that is
responsible for the numerical efficiency of the FLIPT algorithm is
exposed here by expressing the discretized integral as a product of
tensors.  Specifically, frequency indices $k_1 \cdots k_n$ are
expressed as an additional index (superscript $k$ below) which is
summed over.  In this notation, the density matrix at order $n$ from
eq.~(\ref{eq:flipt-main}) is given by,
\begin{align} 
  \rho_n(t) &=  e^{\eta t} \sum_k e^{i\Omega k t} \hat \rho_n^k.\label{eq:hat-rhon-t} 
\end{align}
where $\hat \rho_n^k$ is the $n$-th order frequency-resolved (denoted
by a caret) density matrix at frequency $\Omega k$,
\begin{align}
  \hat \rho_n^k &= \left(\frac{\lambda}{2\pi}\right)^{n}  \sum^\infty_{k_n=\infty}\cdots \sum^\infty_{k_1=\infty} \delta_{k, k_1 + k_2 + \cdots + k_n} \label{eq:hat-rho-n}\\
                   &\times E[k_n] \cdots E[k_2] E_\eta[k_1]\nonumber\\
                         &\times \super G_0\left(i\Omega \sum^n_{i=1} k_i + \eta\right) \super V \super G_0\left(i\Omega \sum^{n-1}_{i=1}k_i+ \eta\right) \cdots\nonumber\\
  & \times \super V\super G_0(i\Omega k_1 + \eta)\super V \rho_0.\nonumber  
\end{align}
The frequency-resolved density matrix $\hat \rho_n$ is then a 3-tensor
with two indices over the quantum mechanical states from the density
matrix and one index over the frequency.  (The parameter $\eta$ is
implicit and is zero for order $n=0$).  The zeroth-order term is
time-independent and thus is nonzero only at the frequency index
$k=0$.  Thus, the initial state has the following tensor form,
\begin{align}
  \hat \rho_0^k = \delta_{k,0} \rho_0.
\end{align}
The next order of the perturbative expansion is given by first applying
$\super V$ to $\rho_0$, then multiplying the result by $E_\eta[k]$ and followed by applying $\super
G_0(i\Omega k + \eta)$.  Each of those operations can be expressed by
a tensor product with a frequency-resolved superoperator.  The tensors
$\hat G$ and $\hat V$ are diagonal in the frequency indices
$k,k'$ and are given by 
\begin{align}
  \hat G^{k',k} &= \delta_{k',k} \super G_0(i\Omega k + \eta)\\
  \hat V^{k',k} &= \delta_{k',k} \super V.
\end{align}
The tensor $\hat E_\eta$, diagonal in the system state indices,
describe the action of $E_\eta(\omega)$ on the frequency of the
system, with
\begin{align}
 \hat E_\eta^{k',k} = \frac{\lambda}{2\pi} E_\eta [k-k']. \label{eq:hatE}
\end{align}
In effect, $\hat E_\eta$ raises or lowers the frequency indices of a
state by $k$ for every nonzero $E_\eta[k]$.  Then, the first order
frequency-resolved density matrix is given succinctly by the following
tensor product,
\begin{align}
  \hat \rho_1 = \hat G \hat E_\eta \hat V \hat \rho_0. 
\end{align}

The iterated structure of the perturbative expansion
[Eqs.~(\ref{eq:rec1}) and (\ref{eq:rec2})] translates naturally into a
repeated tensor product form.  Indeed, higher order terms can be
obtained simply by the repeated application of $\hat G \hat E \hat V$.
In the tensor notation, the order $n+1$ term is given by
\begin{align}
  \hat \rho_{n+1} = \hat G \hat E \hat V \hat \rho_{n}
\end{align}
Note that for $n>1$ 
$E[k]$ (i.e. with $\eta=0$) is used  instead of
$E_\eta[k]$.  Using this notation, the $n$-th order contribution to
the density matrix, from equation (\ref{eq:flipt-main}), is a product
of the tensors defined above,
\begin{align}
  \rho_n(t) = e^{\eta t}\sum_k e^{i\Omega k t} \left[(\hat G \hat E \hat V)^{n-1}\hat G \hat E_\eta \hat V \hat \rho_0 \right]^k\label{eq:tensor-products}
\end{align}
where the square bracket with superscript $k$ denotes the $k$-th
frequency element of the overall product of tensors. The tensor form
described here not only makes for a convenient notation, it is also
much more efficiently evaluated, as shown below.

\subsection{Algorithm}

The tensor structure of eq.~(\ref{eq:tensor-products}) and the uniform
discretization of the Fourier series are exploited in FLIPT method to
reduce the amount of numerical operations performed to compute
eq.~(\ref{eq:flipt-main}).  This is best described by an example, as
done here.  The discretized perturbation $E_\eta[k]$ from
eq.~(\ref{eq:fourier-series}) above is taken to be computed 
at specific values of $\Omega$ and $\eta$ and truncated at frequency
indices $L < |k| < U$.  The number of grid points of the discretized
perturbation is 2$N_d$, where $N_d = U - L$. In this section, we focus
on the frequency integral and thus on the action of the operator $\hat
E$; the frequency-resolved Green's function $\hat G$ and potential
$\hat V$ operators are discussed in Sec.~\ref{sec:tensor-ops}.

The frequency-resolved density matrix at order zero consists of only
one element, $\rho_0$, with frequency index $k=0$.  The first order
contribution is obtained by performing $\hat G \hat E_\eta \hat V \hat
\rho_0$. Specifically, the $k$-th frequency index of $\hat E_\eta \hat
V \hat \rho_0$ is given by the product of $E_\eta[k]$ (a scalar) and
$\super V \rho_0$ (the sole nonzero density matrix of $\hat
\rho_{0}$),
\begin{align}
 \hat \rho_{1'}^k \neq 0 \text{ for } k \in [-U, -L] \text{ and } [L, U].
\end{align}
Then, $\hat \rho_{1} = \hat G \hat \rho_{1'}$ yields the first order
contribution.
Hence, at first order, $2N_d$ density matrices are obtained, with
frequency indices spanning $[L,U]$ and $[-U, -L]$.

The second order $\hat \rho_2$ is obtained from the first order
contribution $\hat \rho_1$ by repeating this process.  First, $\super
V_0$ is applied to each of the $2 N_d$ density matrices of $\hat
\rho_1$ to obtain $\hat \rho_{1''}$.  Then the product with $\hat E$
is performed as follows,
\begin{align}
 \rho_{2'}^{k + k'} = \sum_{k, k'}  E[k'] \rho^k_{1''}.
\end{align}
That is, every nonzero density matrix $\hat \rho_{1''}^k$ is
multiplied with every nonzero values $E[k']$ and summed into $\hat
\rho_{2'}^{k + k'}$.  This yields the following nonzero frequency
indices,
\begin{align}
 \hat \rho_{ 2' }^k \neq 0 \text{ for } k \in [-2U, -2L] , [-N_d, N_d], \text{ and } [2L, 2U].
\end{align}
Thus, $\hat G \rho_{2'} = \hat \rho_{ 2 }$ consists
of 8 $N_d$ terms.

Importantly, there are $2 N_d \times 2 N_d = 4 N_d^2$ possible
products of $E[k']$ and $\rho_{1''}^k$, but only $8 N_d$ distinct
values of $k + k'$.  That is, there are more than one ``pathways'' to
a given frequency index.  For example, the density matrix $\hat
\rho_1^k$ multiplied by $E[-k']$ contributes to the second order
density matrix at index $k - k'$; the term $E[k]\hat \rho_1^{-k'}$
also sums to the same frequency index.  Contributions such as these to the
same final frequency index are summed over
as soon as they are available. \textit{Significantly, this summing is
responsible for the exponential speedup of the FLIPT algorithm over
conventional numerical quadrature.}

Specifically, the number of nonzero frequency indices of $\hat \rho_n$
is linear in $N_d$ as a consequence of the large number of repeated
indices.  The frequency indices where $\hat \rho_n$ is nonzero are
given by,
\begin{align*}
  \hat \rho_0 &\rightarrow [0]\\
 \hat \rho_1 &\rightarrow [-U, -L] \text{ and } [L,U] \\
 \hat \rho_2 &\rightarrow [-2U, -2L] , [-N_d, N_d], \text{ and } [2L,2U] \\
 \hat \rho_3 &\rightarrow [-3U, -3L] , [-U-N_d, -L + N_d]  , [L-N_d , U + N_d], \text{ and } [3L, 3U]\\
              &\cdots 
\end{align*}
At order $n$, there are $n+1$ such intervals, each with $n N_d$
nonzero frequency indices.\footnote{
  The regions are taken here to be disjoint.  When the regions
  overlap, fewer points are needed and the numerical effort is reduced.}
The number of distinct frequency indices
is thus $(n+1) n N_d$, i.e. a linear function of $N_d$.

In contrast, if the frequency discretization is performed using an
arbitrary set of grid points $\{\omega_i\}$, the number of nonzero
elements of $\hat \rho_n$ would scale exponentially with $n$.  Indeed,
given an arbitrary set of frequency points $\{\omega_k\}$ of size
$2N_d$, the set of points $\omega_i + \omega_j$ for all $i$ and $j$
(i.e. the number of frequency indices of $\hat \rho_2$ from a general,
non-uniform grid) contains $ 4 N_d^2$ unique frequencies.  The third
order result (i.e. the number of frequency indices of $\hat \rho_3$)
contains $2 N_d \times 4 N_d^2 = 8 N_d^3$ unique frequencies, the
fourth order result contains $32 N_d^4$ etc. That is, the number of
frequency indices would increase exponentially.  However, the Fourier
series representation (and resulting uniform grid) used here bypasses
the exponential increase without introducing artificial
coarse-graining or filtering of any sort due to the summing of
repeated indices.

\subsection{Numerical analysis}

Formally, the FLIPT algorithm scales as $O(N_d^2)$ at all orders $n$
of the perturbation, instead of the exponential $O(N_d^n)$ scaling of
standard quadrature.  Consider the standard quadrature of
eq.~(\ref{eq:rhon_ss}) using a grid with the $2 N_d$ of points.  The
integrand for the $n$-th perturbative contribution has $n$
applications of both $\super V$ and $\super G_0(i \omega +
\eta)$. Denoting the number of operations required to perform $\super
V \rho$ and $\super G_0(i\omega + \eta)\rho$ by $N_V$ and $N_G$
respectively, the number of operations required to compute the
integrand once is $n (N_v + N_G)$.  The integrand has to be evaluated
at every multidimensional quadrature points.  At order $n$, computing
the $n$-dimensional quadrature using a one-dimensional grid of $2N_d$
points require $(2 N_d)^n$ integrand evaluation.  Thus, the number of
operations required by standard quadrature obeys,
\begin{align}
  N^{(\text{quad})}_n = (2 N_d)^n  n (N_V + N_G) 
\end{align}
Therefore, standard quadrature integration scales as $N_d^n$ ---
exponential in the perturbative order $n$.

In contrast, the FLIPT algorithm presented above requires a number of
operation at most quadratic in the number of discretization points
$N_d$ at all orders.  The equivalent
number of operations required to compute the $n$-th perturbative order
can be derived from eq.~(\ref{eq:tensor-products}) above,
\begin{align}
  N^{(\text{FLIPT})}_n = N^{(\text{FLIPT})}_{n-1} + N_{d,n-1} N_V + N_{d,n} N_{d,n-1} N_E + N_{d,n} N_G.
\end{align}
The first term corresponds to the number of operations required to
obtain the $n-1$ order from which order $n$ is computed. The remaining
terms are the number of operations required to compute tensor products
with $\hat V$, $\hat E$ and $\hat G$ respectively.  In most
applications, the highest order of the perturbative expansion
dominates the CPU time and other terms can be ignored, such that the
following complexity is obtained,
\begin{align}
  N^{(\text{FLIPT})}_n &= O([ N_V + N_G ] N_d) + O(N_d^2 N_E)\label{eq:complexity}.
\end{align}
For a system of $d$ levels, $N_E \propto d^2$, as will be shown below.
Depending on the number of operations required by $\hat V$ and $\hat
G$, the FLIPT algorithm is linear to quadratic in the number of
frequency points $N_d$, an enormous improvement over standard
quadrature for perturbative order 3 and above.

The FLIPT method is well-behaved numerically as the Fourier series
representation is numerically ``exact'' for continuous and time-limited
$E(t)$.\cite{crump_numerical_1976}  Furthermore, the summation
procedure over repeated indices described above is analytic and not
the result of any numerical coarse-graining.  The numerical error of a
function $y(t)$ computed by Fourier-Laplace inversion, such as e.g. a
matrix element of $\rho(t)$ computed using the FLIPT algorithm, obeys
the following relation,\cite{dubner_numerical_1968}
\begin{align}
  \text{error}(t) \le C \exp(\eta (t - T)) \frac{\cosh \eta t} {\sinh \eta T},\label{eq:laplace-error}
\end{align}
where $C = \max y(t)$ over an interval from 0 to $T$.  Hence, the
error increases on the approach to $T$ in proportion to $\eta$.  Error
analysis can be used to compute an optimal value for $\eta$. Following
Ref.~\onlinecite{piessens_algorithm_1984}, $\eta$ is taken here to be
$2\pi/T = \Omega$, which provides a good balance between the Laplace
inversion error and any floating point errors arising from evaluating
the Green's function near poles of $\super L_0$.  For $t<T/2$ the
relative error is $\le 10^{-3}$.  It should be noted that
eq.~(\ref{eq:laplace-error}) is an upper bound; in practice, the
recurrence of the Fourier series at $T$ dwarfs any numerical errors
due to the inversion.

\subsection{Tensor operations} \label{sec:tensor-ops}

The numerical performance of the FLIPT method depends crucially on a
fast computation of the tensor products of
eq.~(\ref{eq:flipt-main}). Below, the numerical implementation and
performance properties of those operations are discussed for a system
of $d$ levels.  It should be noted that the same operations also
determine the performance of other numerical methods; non-perturbative
propagation methods, for example, also rely on the fast evaluation of
$\super V \rho$ and $\super L_0$.

Computing $\hat \rho' = \hat V\hat \rho$ is done by a straightforward
computation of $\super V \rho$ at each frequency index $k$ (the number
of which is denoted $N_k$) where $\hat\rho^k$ is nonzero,
\begin{align} 
  \hat \rho'^k = \frac{1}{i\hbar}\left(\mu \hat \rho^k - \hat \rho^k \mu\right)
\end{align}
Standard matrix-matrix multiplication is used for this
step.  The computational complexity of $\hat V \hat \rho$ is
given by the complexity of performing the $\mu \rho$ matrix-matrix
multiplication ($N_v \propto d^3$ above) multiplied by $N_k$.

Similarly, $\hat \rho' = \hat G \hat \rho$ is evaluated by computing
$\hat \rho'^k = \super G_0(i\Omega k + \eta) \hat \rho^k$ for all values of 
the frequency index $k$. For a closed $d$-level system with
energy eigenvalues $E_i$, $\super G_0(i\omega + \eta) \rho$ is given
by the following analytical formula,\cite{lowdin_operators_1982}
\begin{align}
  [\super G_0(i\omega + \eta)\rho]_{ij} = \frac{\rho_{ij}}{i\omega - i(E_j - E_i)/\hbar + \eta } . \label{eq:superG}
\end{align}
where $\rho_{ij}$ is the $i,j$ matrix element of $\rho$ in the
eigenbasis of $H_0$ of eq.~(\ref{eq:superL}).  Thus, for such a
system, the number of operations per frequency index $N_G = O(d^2)$
and the overall complexity of computing $\hat \rho'$ is $O(N_k d^2)$.

The Green's function can still be evaluated even when $\super L_0$
does not have the form of eq.~(\ref{eq:superL}) or when the
Hamiltonian is not diagonal.  This is the case in,
e.g., Ref.~\onlinecite{lavigne_two-photon_2019} where the Redfield equation
is used.  The Green's function applied to a density matrix $\rho$
yields
\begin{align}
  \rho' = \super G_0(z) \rho = \left[z-\super L_0\right]^{-1} \rho
\end{align}
Multiplying both sides by $(z - \super L_0)$, the following is obtained,
\begin{align}
  \left(z - \super L_0\right) \rho' = \rho.
\end{align}
That is, the product of the Green's function on $\rho$ can be obtained
by solving the above system of linear equations for $\rho'$.  Using an
iterative method (such as GMRES,\cite{baker_technique_2005} used in
Ref.\onlinecite{lavigne_two-photon_2019}) and an efficient algorithm
for the product $\super L_0 \rho$ (such as the Pollard-Friesner
method\cite{pollard_solution_1994}), a numerical complexity of
$O(d^3)$ is obtained.  For small systems, direct methods (such as the
LU decomposition\cite{anderson_lapack_1999}) can also be used.

Finally, $\hat \rho' = \hat E \hat \rho$ is computed using a
matrix-matrix product over frequency indices,
\begin{align}
  \hat \rho'^{k'} = \sum^{N_k}_{k}\hat E^{k',k} \hat \rho^{k}.
\end{align}
where $\hat E^{k',k}$ is given by eq.~(\ref{eq:hatE}).  This operation
has a numerical complexity of $O(N_k' N_k d^2)$ where $N_k$ and $N_k'$
are the number of indices $k$ and $k'$ where $\hat \rho^k$ and $\hat
\rho'^{k'}$ respectively are nonzero.  This is done numerically by a
matrix-matrix multiplication with $\hat E$ expressed as a $N_k' \times
N_k$ matrix and $\hat \rho$ expressed as a $N_k \times d^2 $ matrix.

Importantly, it should be noted that the quadratic scaling described
in this paper is for the case where the computation of $\hat E$ is the
most expensive step.  This is true when the number of levels is low
and $\super V \rho$ and $\super G_0(\omega)\rho$ are relatively
inexpensive.  For larger systems, this is no longer the case and the
FLIPT algorithm becomes dominated by the linear in $N_d$ terms of
eq.~(\ref{eq:complexity}); in those cases, the performance advantage
of FLIPT over other quadrature method is significant even for second
order processes such as linear absorption.

\section{Implementation and example calculations}
The FLIPT algorithm is easily implemented using tensor algebra.  Such
an implementation for closed and open secular dynamics is made freely
available by the authors.\cite{lavigne_flipt.jl_2019}  The resultant
code is very short (less than 600 lines in the Julia programming
language) and can thus be easily translated to other programming
languages and environments.

The implementation can be used as a ``black box'' to compute arbitrary
orders of the perturbative expansion as well as arbitrarily complex
nonlinear spectroscopy diagrams.  The implementation includes
subroutines to build $\hat V$ and $\hat G$ from $\mu$ and $H_0$ for
closed systems.  The ability to optionally operate on only the ket or
bra sides of the density matrix with either solely positive or
negative frequency components of the field can be used to compute the
response of arbitrary nonlinear spectroscopy diagrams including $\vec
k$-vector phase matching.\cite{mukamel_principles_1995}  The maximum
propagation time $T$, the energy levels, the dipole transition
operator and the electric field $E(\omega)$ are the only required
parameters; the implementation is otherwise fully automatic.

We have recently applied the FLIPT algorithm to the simulation of
quantum dynamics arising from two-photon absorption in retinal to
understand the mechanism of a quantum control experiment on living
brain cells.\cite{paul_coherent_2017,lavigne_two-photon_2019}  Here,
we focus on the performance properties of the FLIPT algorithm, and the
application of the FLIPT method to spectroscopy.  First, performance
is studied using a small model system; results are compared with
non-perturbative propagation.  Then, a sample calculation on a model
molecule, pyrazine, shows how the FLIPT method can be applied to the
simulation of spectroscopy experiments.  The model exhibits an
ultrafast conical intersection mediated population transfer between
two electronic states.  A signature of the conical intersection is
present in a simulated transient absorption spectrum obtained from a
third-order perturbative calculation.

\subsection{Numerical scaling and performance}
\begin{figure}[h]
  \begin{subfigure}[b]{0.4\textwidth}
    \input{lambda.tex}
    \caption{\label{fig:lambda-model}}
  \end{subfigure}
  \begin{subfigure}[b]{0.4\textwidth}
    \includegraphics[width=\textwidth]{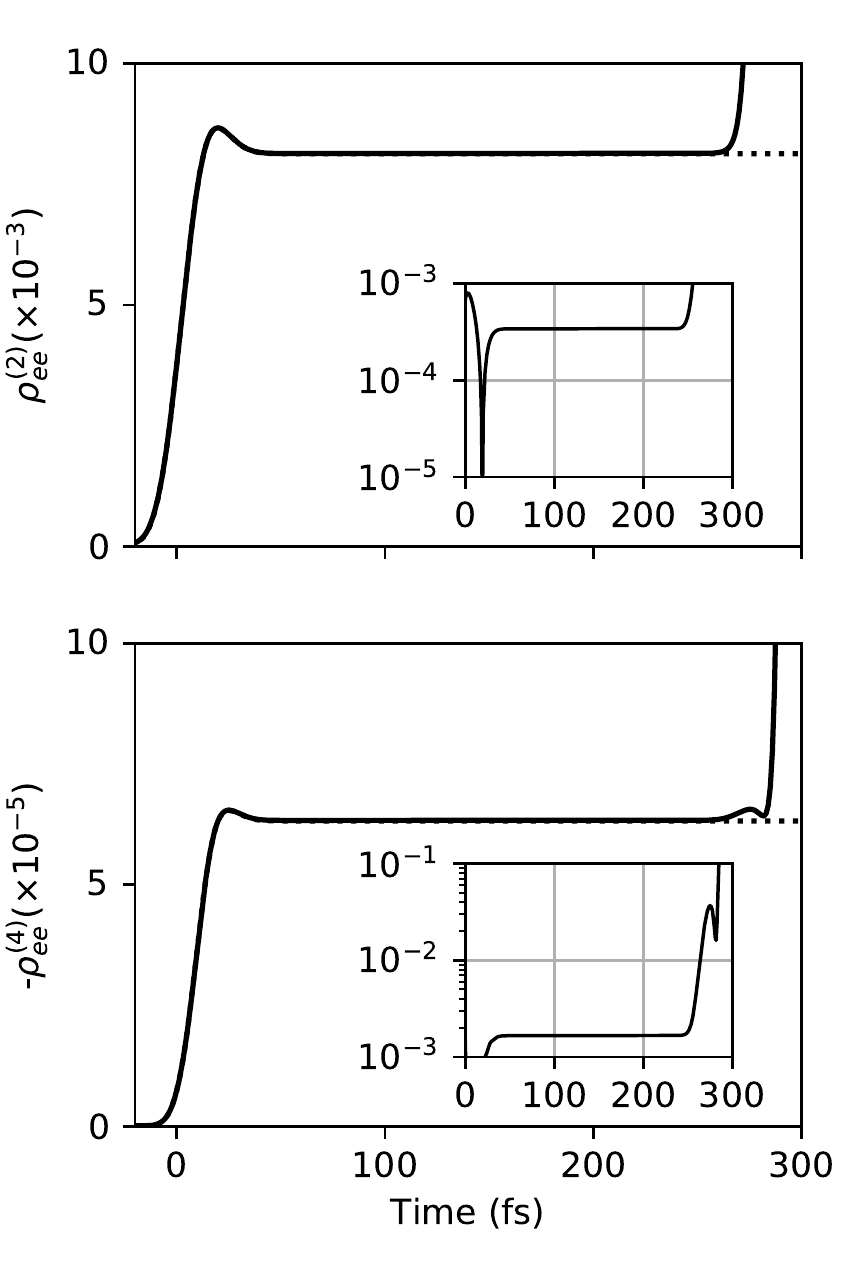}
    \caption{\label{fig:lambda-direct}}
  \end{subfigure}
  \caption{
    (a) Model system used to compare the FLIPT method with
    non-perturbative propagation.  The states $\ket{g}$, $\ket{e_1}$,
    $\ket{e_2}$ and $\ket{g'}$ have energy 0, 1.95, 2.05 and 0.05 eV
    respectively.  The dipole transition matrix elements are $\mu_{g,e_1}
    = 0.1$, $\mu_{g,e_2}=0.2$ $\mu_{g',e1} = -0.25$ and $\mu_{g',e_2} =
    0.15$ eV/$E_0$.  (b) Comparison of the FLIPT (solid) and
    non-perturbative (dashed) population of the $\ket{e_i}$ manifold.  The
    top and bottom figures show linear and quadratic contributions in the
    intensity of the field, which is a Gaussian pulse with a FWHM of 30 fs
    and a central frequency of $2.0 $ eV$/\hbar$.  The propagation time
    $T$ is 300 fs in either case.  The inset shows the relative error of
    the FLIPT result.  Note difference in ordinate scale in Figs 2(a) and
    2(b). }
  \label{fig:model-params}
\end{figure}

The numerical performance and numerical error of the FLIPT method
are studied here using a four-level model system
(Fig.~\ref{fig:lambda-model}).  This small model exhibits slow
($\approx 100 $ fs) and fast dynamics ($\approx 1 fs$) dynamics
resulting from separated manifolds $\ket{g}$ and $\ket{e}$, each
consisting of two closely spaced levels.  The system interacts with an
ultrafast, coherent pulse.  The spectrum of this pulse is given by,
\begin{align}
  E(\omega) &= g(\omega; \omega_0, \sigma) + g(\omega; -\omega_0, \sigma),\label{eq:gaussian1}
\end{align}
where the function $g(\omega; \omega_0, \sigma)$ is a dimensionless,
normalized Gaussian function centered at frequency $\omega_0$,
\begin{align}
  g(\omega; \omega_0, \sigma, \chi) &= \frac{1}{(2\sigma)^{1/2}\pi^{1/4} }
  \exp\left(-\frac{(\omega-\omega_0)^2}{2\sigma^2}\right).
\end{align}
The standard deviation of the field $\sigma$ in the Fourier domain is
obtained from the FWHM in the time domain as,
\begin{align}
  \sigma = \frac{2\sqrt{2\log 2}}{\text{FWHM}_t}.\label{eq:gaussian2}
\end{align}
The FLIPT results are compared to the result of a non-perturbative
propagation of the density matrix.  A perturbative decomposition of
the non-perturbative result is approximated with a least-squares fit
at multiple values of the field intensity.  Computations are performed
using one core of an Intel Xeon E5-2630 (2.20GHz) processor, with
parameters given in the caption of Fig.~\ref{fig:model-params}.

It should be noted that no artificial decoherence or broadening
parameters are introduced --- the only parameter of the FLIPT method
is the propagation time $T$.  The value of $T$ sets the maximum
duration of a valid FLIPT result; indeed, the Fourier series
decomposition imply a recurrence of the dynamics at time $T$ after the
excitation.  Away from the recurrence at $T$, the FLIPT method is
highly accurate even when compared to the non-perturbative
method, as shown in Fig.~\ref{fig:lambda-direct}.  In the quadratic
case, the relative error between the FLIPT result and non-perturbative
propagation is less than $10^{-3}$ and stable until at least $T/2$.
The quartic case is similar; the higher relative error is due to the
order extrapolation of the non-perturbative
propagation.

\begin{figure}[h]
  \includegraphics[width=0.5\textwidth]{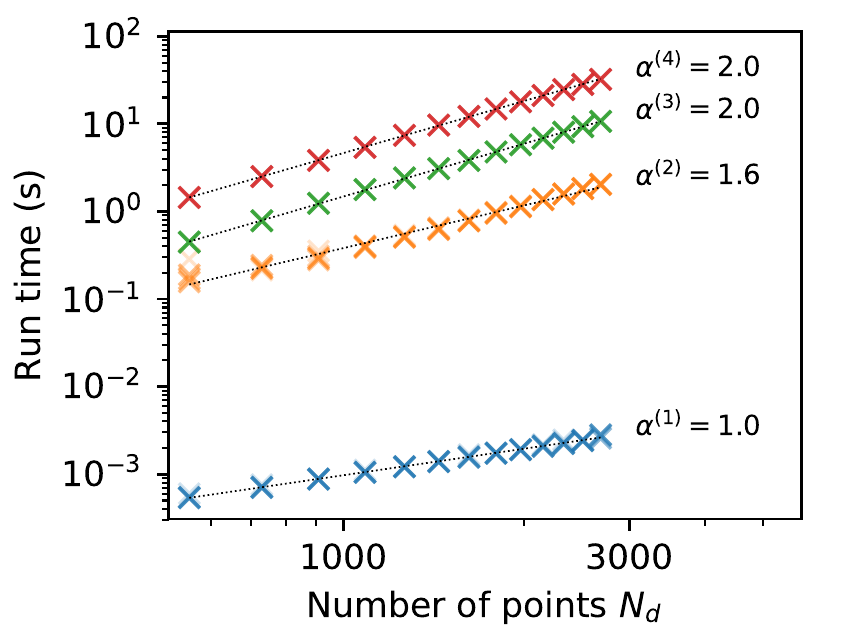}
  \caption{
    Numerical scaling $O(N_d^\alpha)$ of the FLIPT method with respect
    to the number of frequency points $N_d$ for each order $n$ of a
    fourth-order perturbative expansion of the model of
    Fig.~\ref{fig:lambda-model}.  Dotted lines show a least-squares
    fit for the scaling parameter $\alpha^{(n)}$ at order $n$.
  }
  \label{fig:timingspt} 
\end{figure}

The linear to quadratic scaling with respect to the number of
frequency points given in eq.~(\ref{eq:complexity}) is demonstrated in
Fig.~\ref{fig:timingspt}, where the duration $T$ (and thus the number
of points $N_d$) of the propagation is adjusted. The time required to
compute each of the first four perturbative orders of $\rho(t)$ are
shown.  As described above, the complexity is at most quadratic for
all perturbative orders, making the FLIPT method highly suited to the
computation of higher-order processes.

\begin{figure}[h]
  \includegraphics[width=0.5\textwidth]{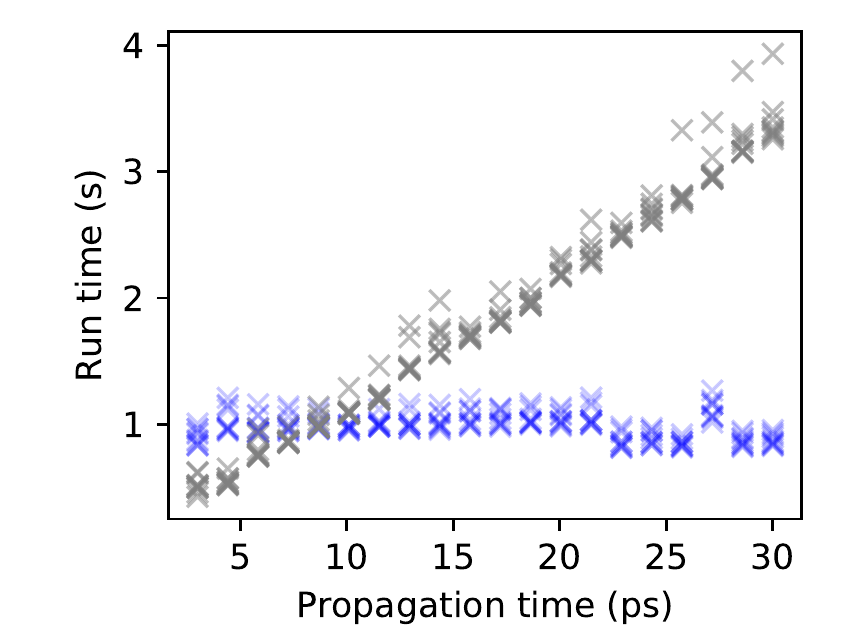}
  \caption{
    Computation time of a non-perturbative (gray) and FLIPT (blue) propagation
    of the $\Lambda$ model of Fig.~\ref{fig:lambda-model} after excitation
    with Gaussian pulse with a FWHM of $T/10$ where $T$ is the maximum
    propagation time. \label{fig:timings}
  }
\end{figure}

The FLIPT method is uniquely suited to study processes where fast and
slow timescales (e.g., electronic and nuclear motion) are both
present. The spectral nature of the method leads to an important
property: time-scale invariance.  This sets it apart from
non-perturbative propagation and other time-dependent
methods.\cite{rose_numerical_2019,
arango_communication_2013,zhuang_simulation_2006} Indeed, the FLIPT
method is independent of the energy spacing of the system under
investigation, since that spacing is fully subsumed in the smooth
function $\super G_0(\omega + i\eta)$.  The only timescales of
importance are those of the interaction and of the duration of
propagation $T$.  The number of points $N_d$ required is directly
proportional to the bandwidth of $E(\omega)$ and inversely
proportional to the duration of the simulation, a result similar to
that obtained with the rotating wave approximation, but here without
any approximations.  Hence, computing dynamics over 300 ps arising
from a 10 ps interaction with frequency $\omega_0$ requires the same
computation time as computing 300 fs of dynamics arising from a 10 fs
pulse.  In contrast, a non-perturbative propagation of these cases
will scale at least linearly as the minimal time step is set by the
fast, optical transients of the electric field.  This property is
demonstrated in Fig.~\ref{fig:timings}.

\subsection{Sample calculation: spectroscopy of pyrazine}
Pyrazine, a small molecule, is a well-known model system to study ultrafast
internal conversion.\cite{
  raab_molecular_1999,sukharev_optimal_2004,
  christopher_overlapping_2005,christopher_quantum_2006,
  ryabinkin_when_2014}
A model for pyrazine with three electronic surfaces (denoted $S_0$,
$S_1$ and $S_2$) and two vibrational modes (denoted $6a$ and $10a$) is
used here to demonstrate the use of FLIPT with a sizable multilevel
system (with $\approx 300$ levels).  The Hamiltonian for this system
is given by,
\begin{align}
  H &= 
  \begin{pmatrix}
    \epsilon_0 + h_0 & 0 & 0\\ 
    0 & \epsilon_1 + h_0 +  \kappa_1 q_{6a} & \lambda q_{10a}\\ 
    0 & \lambda q_{10a}  & \epsilon_2 h_0 + \kappa_2 q_{6a} 
  \end{pmatrix}\\
  h_0 &= \sum_{i \in 6a, 10a} \frac{\omega_i}{2} \left(p_i^2 + q_i^2\right).
\end{align}
The operators $p_i$ and $q_i$ are the momentum and position operators
for the modes in mass-weighted coordinates, $\epsilon_i$ is the Franck-Condon
energy of surface $i$, $\kappa_i$ is the displacement of the tuning
mode $6a$ on surface $i$ and $\lambda$ is the non-adiabatic coupling
between surfaces $S_1$ and $S_2$.  The transition dipole operator $\mu$
couples the ground and excited surfaces,
\begin{align}
  \mu = E_0 \mu_0 \begin{pmatrix}
    0 & \sqrt{0.2} & 1 \\
    \sqrt{0.2} & 0 & 0 \\
    1 & 0 & 0
  \end{pmatrix},
\end{align}        
where $E_0\mu_0=1 eV$ is a scaling parameter for the perturbative
expansion.  Parameters for this model are given in
Table~\ref{tab:params}.  Direct diagonalization yields $\approx 300$
basis states.  Gaussian pulses, described by
eqs.~(\ref{eq:gaussian1})-(\ref{eq:gaussian2}), are used throughout.

\begin{table}
  \begin{ruledtabular}
    \begin{tabular}{ccccccc}
      $\omega_t$ &$\omega_c$ &$\kappa_1$ &$\kappa_2$ &$\epsilon_1$ &$\epsilon_2$ &$\lambda$ \\
      \hline
      0.0739 &0.1139 &-0.0981 &0.1355 &3.94 &4.89&0.1830 \\
    \end{tabular}
  \end{ruledtabular}
  \caption{
    Parameters used for the 2D models of pyrazine obtained from
    Ref.~\onlinecite{sukharev_optimal_2004}.  All values have units of eV.}
  \label{tab:params}
\end{table}

\begin{figure}[h]
  \includegraphics[width=\textwidth]{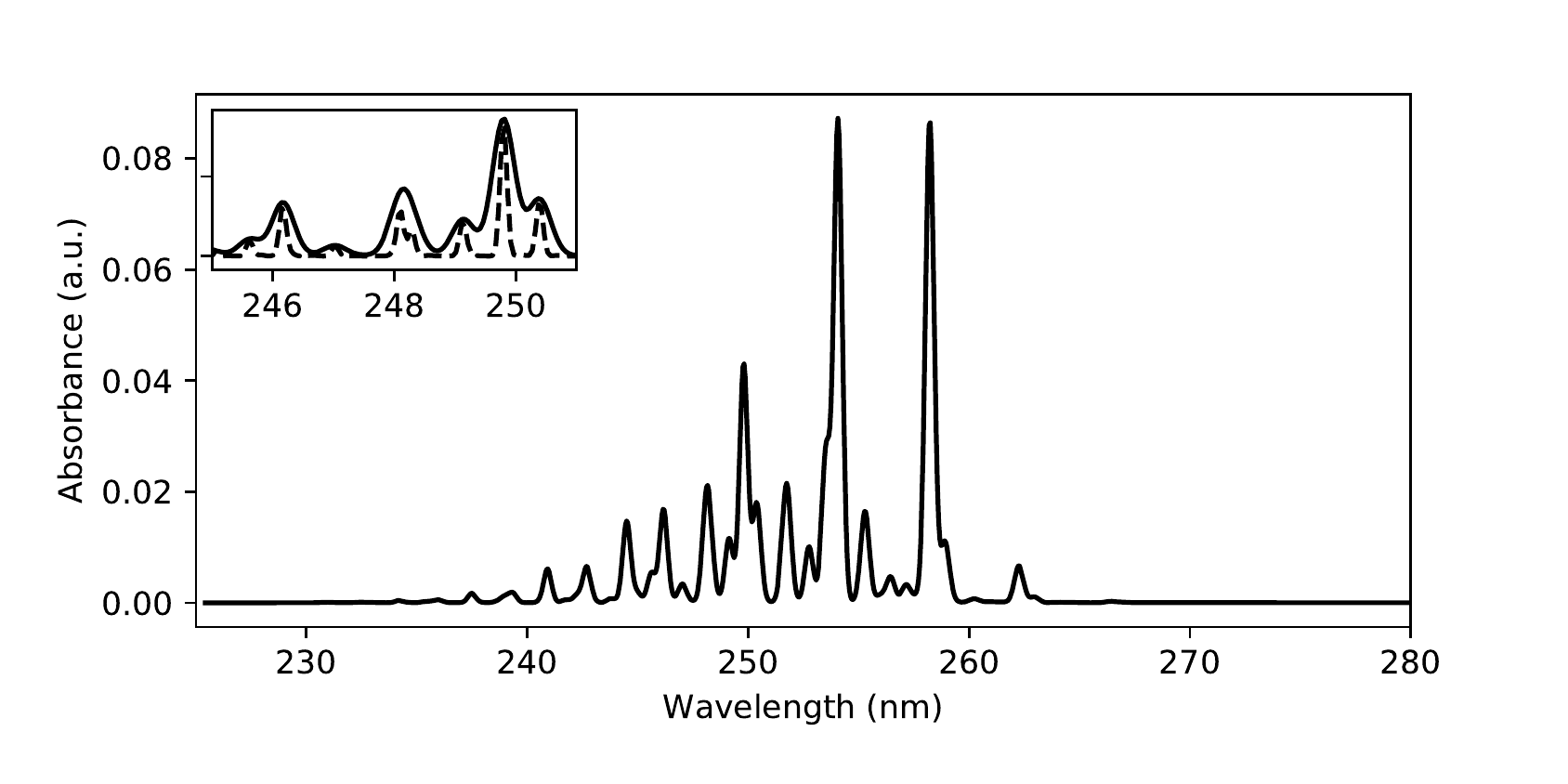}
  \caption{
    Calculation of the absorbance of pyrazine using Gaussian pulses
    with a FWHM of 0.3 ps.  The inset shows the gain in resolution obtained
    using longer pulses, with a FWHM of 1 ps (dashed).}
  \label{fig:pyrlinear}
\end{figure}

A measurement of the linear absorption is simulated using the FLIPT
algorithm (Fig.~\ref{fig:pyrlinear}). The absorption is computed
from the heterodyne detection formula given by
eq.~(\ref{eq:heterodyne-integ}) in the Appendix,
\begin{align}
  I_\text{het}(\omega_0) = I_\text{out}(\omega_0) - I_\text{in}(\omega_0) \propto   \int_{-\infty}^{\infty}\mathrm d\omega E^*(\omega) \mu(\omega),
\end{align} 
where $\omega_0$ is the central frequency of the field $E(\omega)$, a
Gaussian pulse with a FWHM of 300 fs.  The central frequency of the
pulse is swept to obtain the absorbance,
\begin{align}
  A(\omega_0) = -\log_{10}\left[1 + I_\text{het}(\omega_0)/I_\text{in}(\omega_0) \right].
\end{align}
The resolution of the absorbance spectrum is directly proportional to
the length of the (minimum uncertainty) probe pulses.  This is shown in
the inset of Fig.~\ref{fig:pyrlinear}, where the absorbance spectrum
obtained with pulses with a FWHM of 1000 fs is compared with that
obtained with the above shorter pulses.  Using pulses that are longer
in the time-domain and narrower in the frequency domain significantly
increases the resolution of the spectral peaks, a finite pulse
effect.\cite{smallwood_analytical_2017,perlik_finite_2017}  It should
be noted that both results are computed using the same number of
discrete frequency points per pulse and thus require the same amount
of computing time.  Furthermore, no additional decoherence processes,
phenomenological broadening or system-bath interactions are added: the
``smoothness'' of the spectrum is entirely due to the limited
resolution of the probe laser.  This should be contrasted with other
common methods of computing the spectrum, e.g., through the Fourier
transform of the autocorrelation function\cite{raab_molecular_1999} or
the response of the system to a CW field, where \textit{ad-hoc}
broadening factors or signal windowing are required to obtain
numerical convergence.

\begin{figure}[h]
    \includegraphics[width=0.5\textwidth]{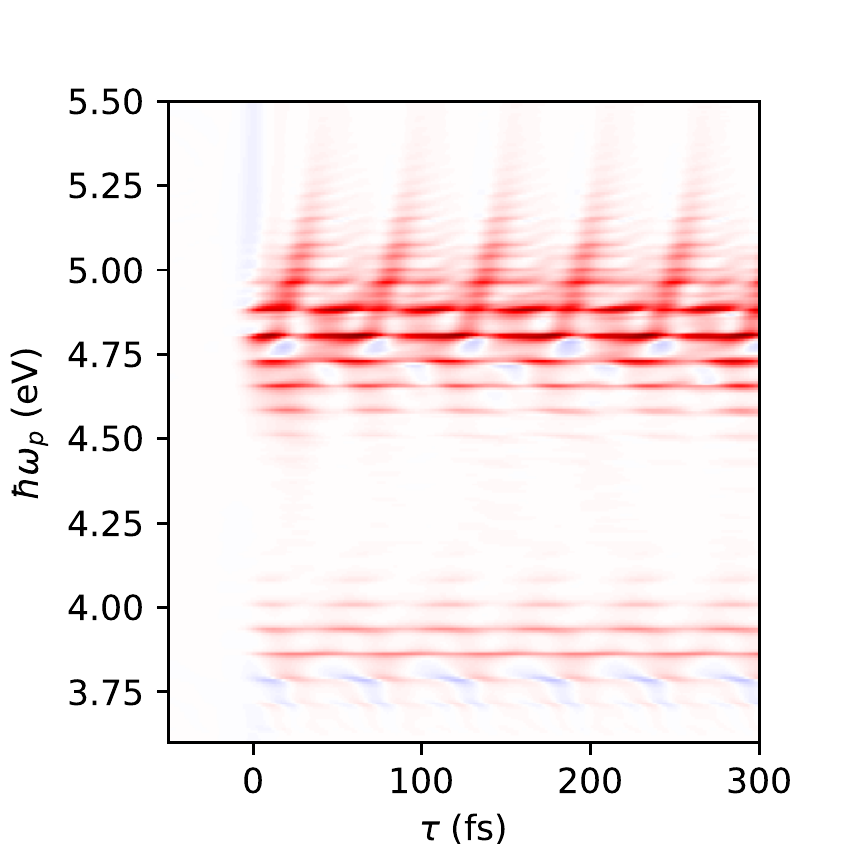}
    \caption{Transient absorption spectrum [Eq.~(\ref{eq:TA})] for the
      pyrazine model as a function of the pump-probe separation time $\tau$ and
      absorption energy $\hbar \omega$. Red and blue regions correspond to increased and decreased
      absorbance, i.e. to positive and negative values of eq.~(\ref{eq:TA}).
      \label{fig:pyrTA} }
\end{figure}

A pump-probe transient absorption (TA) spectrum is computed as an
example of the kind of higher order process that can be studied with
the FLIPT method.  In a transient absorption experiment, a linear
absorption spectra is measure with an ultrashort probe pulse after
excitation with an ultrashort pump pulse and plotted as a function of
the delay $\tau$ between the pump and probe pulses. This is a modeled
four-wave mixing experiment and thus third order in the perturbative
series; a three-dimensional integral is computed at every value of the
pump-probe delay.

Here neither the pump nor the probe are approximated; a realistic
pulse shape is used in both case.  The pump pulse has a FWHM of 20 fs
and is centered at 4.8 eV, the absorption maximum of $S_2$.  The probe
pulse $E_\text{probe}(\omega)$ has a FWHM of 5 fs and a central
frequency of 4.3 eV; its bandwidth is sufficient to probe both the
$S_1$ manifold between 3.7 and 4.2 eV and the $S_1$ manifold between
4.5 and 5.5 eV. This simulation require in this case a discretization
of 180 points (positive and negative) for the pump and 360 points
(positive only) for the probe.  Using the same discretization with a
standard quadrature calculation would require more than 10 million
integrand evaluations, each computed using seven matrix-matrix
multiplications of $\mu$ and $\rho$.  In comparison, the FLIPT result
is evaluated using only 1000 such multiplications --- a factor of
$10^4$ improvement. This integral is repeated for 300 different values
of the pump-probe delay.

The absorbance measured with the probe pulse is given by,
\begin{align}
  A_\text{probe}(\omega_p) &= -\log_{10}\left[1 + I_\text{het,probe}(\omega_p)/I_\text{probe}(\omega_p) \right],
\end{align}
where the heterodyne intensity is computed from
eq.~(\ref{eq:heterodyne}) and the probe intensity is given by
$I_\text{probe}(\omega_p) = |E_\text{probe}(\omega_p)|^2$.  The
absorbance measured with the probe pulse after excitation with the
pump pulse is given by,
\begin{align}
  A_\text{pump-probe}(\omega_p; \tau) = -\log_{10}\left[1 + (I_\text{het,pump-probe}(\omega_p; \tau) +  I_\text{het,probe}(\omega_p))/ I_\text{probe}(\omega_p) \right],
\end{align}
where $I_\text{het,pp}(\omega_p; \tau)$ is the heterodyne-detected
absorption of the probe pulse following excitation with the pump
pulse,\footnote{Due to the strict time-ordering of the perturbative
expansion, $I_\text{het,pp}(\omega; \tau)$ is zero when the probe
arrives before the pump, i.e. it is background-free.} a fourth order
perturbative term.  The transient absorption spectrum is the
difference of these two quantities,
\begin{align}
  \text{TA}(\omega_p,\tau) = A_\text{pump-probe}(\omega_p; \tau) - A_\text{probe}(\omega_p).\label{eq:TA}
\end{align}

\begin{figure}[h]
  \begin{subfigure}[b]{0.45\textwidth}
    \includegraphics[width=\textwidth]{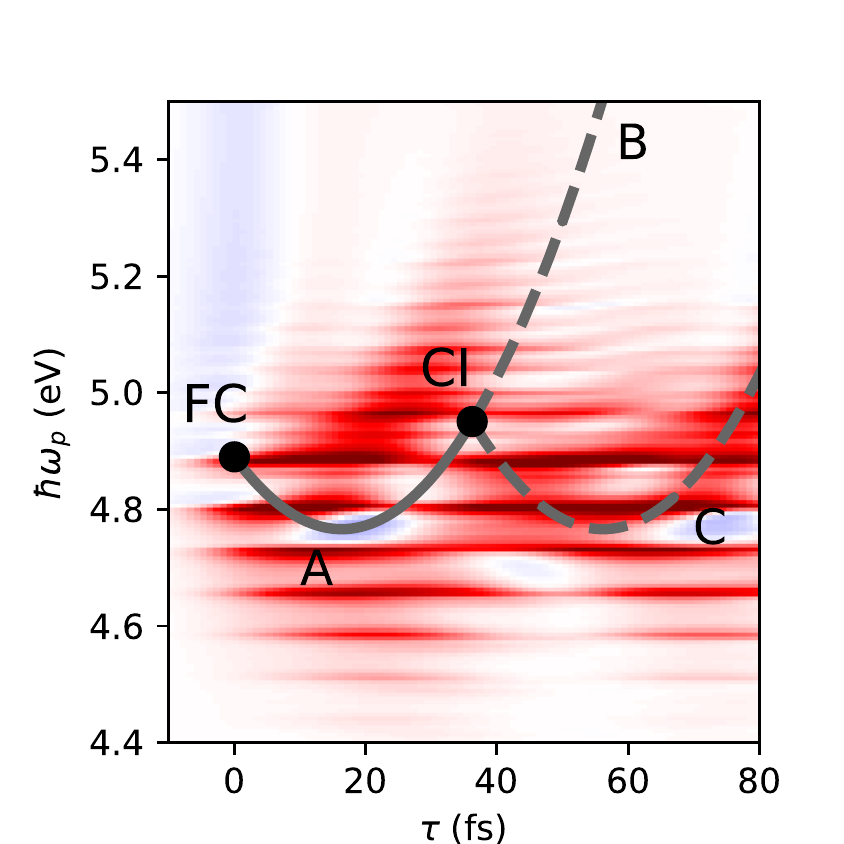}
    \caption{\label{fig:pyrTA-ci}}
  \end{subfigure}
  \begin{subfigure}[b]{0.45\textwidth}
  \input{pump-probe.tex}
    \caption{\label{fig:pyrTA-ci-diag}}
  \end{subfigure}
  \caption{
    (a) Detail of Fig.~\ref{fig:pyrTA} for the first 80 fs and at
    energies corresponding to the $S_2$ surface.  Gray lines show the
    potential energy of a particle (see text) evolving on $S_2$ along the
    tuning mode before (solid) and after (dashed) encountering the conical
    intersection. Segments \textsf{A}, \textsf{B} and \textsf{C} are
    pictured diagrammatically in (b).
  }
\end{figure}

A simulated TA spectrum is shown in Fig.~\ref{fig:pyrTA}.  Red and
blue colors represented heightened and lowered absorption compared to
the unpumped system. The pump pulse excites a coherent wavepacket on
the $S_1$ and $S_2$ surfaces; wavepacket dynamics on those surfaces
generate the time-dependent transient absorption. Fig.~\ref{fig:pyrTA}
shows the initial excitation at delay zero, followed by wavepacket
oscillations on the $S_2$ surface at $\hbar\omega_p\approx$ 4.7 eV and
on the $S_1$ surface at $\hbar\omega_p\approx$ 3.8 eV due to the
tuning mode $6a$. The structured bands in the spectrum are due to
vibronic states of the two electronic surfaces.  As the model includes
no decoherence, there is no decay of the signal.

The TA spectrum reveals details of the two excited electronic surface,
including the presence of a conical intersection.  The first 80 fs of
the TA spectrum in the energy region of the $S_2$ surface are
highlighted in Fig.~\ref{fig:pyrTA-ci}.  Qualitatively, a band of
increasing energy (along the solid curve marked \textsf{B}) can be
identified that splits off at $\tau\approx 40$ fs from the main energy
oscillations around 4.8 eV (along dashed curves marked \textsf{A} and
\textsf{C}).

These two features are produced by the wavepacket motion along the
tuning mode $6a$, shown diagrammatically in
Fig.~\ref{fig:pyrTA-ci-diag}.  Initial excitation by the pump
generates a coherent wavepacket at the Franck-Condon (FC) point. Then,
the wavepacket moves on the $S_2$ surface towards the conical
intersection (CI) between the $S_1$ and $S_2$ surface along the path
marked by \textsf{A} in both figures.  Upon encountering the CI, the
wavepacket bifurcates.  Part of the wavepacket moves past the CI and
continues along the $S_2$ surface, producing the energy increasing
feature along \textsf{B}, while some of the wavepacket transits to
$S_1$ and reflects back onto $S_2$, which yields the energy
oscillation shown by \textsf{C}.  Finally, some of the population
transits onto $S_1$ and moves to the $S_1$ minimum. This latter
contribution is mostly dark at those energies, but is responsible for
the transient absorption signal below the $S_2$ minimum.

The paths \textsf{A}, \textsf{B} and \textsf{C} in
Fig.~\ref{fig:pyrTA-ci} are obtained from the $S_2$ potential energy
$V_2(q(\tau))$ as a function of the normal mode $q(\tau) =
\omega\tau$, where $q(0) = 0$ is the FC point.  That is, the paths are
approximations of the potential energy of a ballistic particle moving
on $S_2$.  Specifically, this qualitative treatment shows how a
conical intersection could be experimentally identified from the
bifurcation of a wavepacket as measured in a transient absorption
experiment. Importantly, all parameters leading to this result have
clear physical origins; the fully automatic FLIPT computation requires
no broadening or convergence factors not directly related to physical
properties of the molecule and radiation.

\section{Conclusion}

Time-dependent perturbation theory plays a central role in
applications of quantum mechanics. However, higher order perturbative
contributions are difficult to evaluate numerically.  Inversion of the
Laplace transform by Fourier
series\cite{dubner_numerical_1968,crump_numerical_1976,piessens_algorithm_1984}
is used to compute arbitrary order time-dependent perturbative
expansions of the Liouville-von Neumann equation with analytically
demonstrated convergence for perturbations of finite
duration. Importantly, nonlinear
spectroscopy\cite{cohen-tannoudji_atom-photon_1992,mukamel_principles_1995}
and quantum control experiments\cite{shapiro_quantum_2012} performed
with pulsed lasers are well described by perturbation theory with
finite duration fields.  Here we have introduced the FLIPT algorithm
that uses the particular structure of the perturbative expansion to
efficiently compute the Fourier-Laplace inversion.  It is a
numerically exact scheme to compute the Fourier-Laplace inversion of
the perturbative series, that has significant advantages over other
integration methods and over non-perturbative propagation methods.
Compared with a propagation of the density matrix, the FLIPT method
yields perturbative results that are more readily understood and
experimentally applicable.  Since it is a spectral method, its
computational complexity is independent of the fast and slow inherent
timescales of the system.  In contrast, time-dependent propagation
methods require fine timesteps when the RWA breaks
down,\cite{arango_communication_2013,rose_numerical_2019} as is the
case in the presence of significant nonresonant processes.

The iterated structure of the perturbative expansion can be expressed
succinctly using tensor algebra; exploiting this structure, the FLIPT
method achieves an exponential speedup with respect to the
perturbation order over standard multidimensional
quadrature.\cite{novak_curse_1997}  Indeed, the computational
complexity of the algorithm is at most quadratic at any order $n$ of
the perturbative expansion, i.e. the $N$-point discretized
multidimensional integration can be computed in $O(N^\alpha)$
operation with $\alpha \lesssim 2$. That is significantly better than
the $O(N_d^n)$ scaling of standard quadrature methods. Furthermore,
the obtained fixed-grid spectral representation is easily Fourier
transformed to the time-domain, a property not shared by quadrature
and Monte Carlo integration.

Other methods for simulating the interaction of light with matter
include the SPECTRON program\cite{zhuang_simulation_2006} and the NISE
method,\cite{torii_effects_2006} as well as symbolic or analytical
approaches.\cite{perlik_finite_2017,smallwood_analytical_2017}
SPECTRON and NISE are significantly less general than the FLIPT
method, since they are limited to computing responses of specific
types of linear to quartic order spectroscopies.  Analytical
approaches are limited to specific idealized pulse shapes and must be
re-derived, with significant effort, for every perturbation order.  In
contrast, the FLIPT algorithm can be used to compute any observable
quantities (i.e. not only responses but also electronic populations,
vibrational displacements, etc.), at arbitrary order of the
perturbation theory and for arbitrary pulse shapes.

The recently proposed UFF method of
Ref.~\onlinecite{rose_numerical_2019} is in many respects similar to
the FLIPT algorithm.  UFF is an arbitrary order, arbitrary interaction
time-domain approach developed for wavefunctions.  Hence, UFF scales
significantly better than the FLIPT method with respect to the size of
the perturbed system.  However, it has convergence issues with respect
to fast oscillations in the absence of the RWA and is limited to
energy-diagonalized Hamiltonian systems.  The choice between the UFF
and FLIPT algorithm should be made based on considerations such as the
size of the system under study, the importance of nonresonant
processes and the order of the perturbative expansion.

Future work should focus on extending the applicability of the FLIPT
algorithm.  Here, the high performance of the method is achieved at
the cost of a high memory usage.  Indeed, for large systems, a copy of
the density matrix must be stored at each discretized frequency index,
with a correspondingly large memory cost.  The use of non-Markovian
equations can significantly reduce memory usage, as can the
propagation of wavefunctions instead of density matrices.  Such
extensions are under development.  Finally, although the primary focus
of this paper has been on light-matter interaction, the FLIPT method
can conceivably be applied to other kinds of perturbation theory.  In
particular, it is closely related to some path integral methods with
similar recursive integral
structures.\cite{thirumalai_iterative_1983,shao_iterative_2001,jadhao_iterative_2008}
Applications to other systems are being investigated.

\textbf{Acknowledgments:} This work was supported by the
U.S. Air Force Office of Scientific Research under Contract No.
FA9550-17-1-0310, and by the
Natural Sciences and Engineering Research Council of Canada.

\bibliography{nl_method}

\appendix
\section{Spectral quantities from FLIPT result}
The Fourier-Laplace inversion can be used to compute spectrally
resolved quantities without introducing \textit{ad-hoc} broadening or
decoherence factors, but care must be taken to ensure convergence.
Indeed, consider the expectation value of an operator $O$ evaluated
using the frequency-resolved, $n$-th order density matrix at index
$k$ [eq.~(\ref{eq:hat-rho-n}) above],
\begin{align}
  \hat O_n^k  = \Tr [O \hat \rho_n^k]
\end{align}
This quantity is not the Fourier transform of $O_n(t) = \Tr O
\rho_n(t)$ at the frequency $\Omega k $.  Specifically, $\hat O_n^k$
is given by a convergent Fourier series over a finite interval while
the Fourier transform of $O_n(t)$ may not even exist.  For example, in
the absence of an environment, oscillatory coherences do not decay and
the Fourier transform of an oscillatory expectation value does not
converge.\cite{lavigne_interfering_2017}  However, approximate Fourier
transforms can be computed using the FLIPT method as shown below.

First, the spectrum of an observable $O$ at perturbative order $n$ can
be approximated by taking the Fourier transform of the time-dependent
value $O_n(t)$ over a finite interval.  The obtained spectrally
resolved expectation value $O_{T,n}(\omega)$, defined below, is an
approximation to the true Fourier
transform.\cite{wiener_generalized_1930}  Using the Fast Fourier
Transform to compute $O_{T,n}(\omega)$ from $O_n(t)$ is expensive in
the presence of high-frequency components since it requires $O_n(t)$
to be meshed over a fine grid.  Fortunately, $O_{T,n}(\omega)$ can be
computed directly from the FLIPT result at the grid points
$\omega=\Omega k$ using the Fourier series,
\begin{align}
  O_{T,n}(\Omega k) &= \frac{1}{T}\int_{0}^{T}\mathrm d t e^{-i \Omega k t} \Tr[O \rho_n(t)]\\
  &=\frac{1}{2 \pi i} \sum_{k'=-\infty}^{\infty}  \frac{e^{\eta T} - 1}{ 2\pi (k' - k) - i\eta T}\Tr O \hat \rho_n^{k'}.
\end{align}
The expectation value $O_{T,n}(\Omega k)$ converges as a distribution
to the Fourier transform when $T\rightarrow
\infty$.\cite{rudin_real_1987}  For the case where $\eta = \Omega$, as
is done in the present implementation, the above further simplifies
to,
\begin{align}
  O_{T,n}(\Omega k) = \frac{e^{2 \pi} - 1}{2 \pi i} \sum_{k=-\infty}^{\infty}  \frac{\Tr O \hat \rho_n^{k'}}{ 2\pi (k' - k - i)}.
\end{align}

Second, specific formulas can be obtained for spectroscopic signals
detected through heterodyning.  This is the case in many ultrafast
spectroscopy experiments, e.g., transient absorption, pump-probe and
2D spectroscopy.  A heterodyne signal $I^{(n)}_\text{het}(t)$ is
obtained by mixing the response of the system, given by an observable
$\mu_n(t)$, with a probe electric field $E(t)$ and detected in the
direction of the probe,\cite{mukamel_principles_1995}
\begin{align}
  I^{(n)}_\text{het}(t) &= E^*_+(t) \mu_{n+}(t),
\end{align}
where the subscripts $+$ denote that only positive frequency
components (i.e. positive phase-matched $\vec k$ components) are
detected.  The heterodyne signal can be computed from the Fourier
series representation as follows,
\begin{align}
  I^{(n)}_\text{het}(t) &= \sum_{k'=0}^{\infty} E^*_{k'} e^{-i \Omega  k' t} \frac{e^{\eta t}}{2\pi} \sum_{k=0}^{\infty} e^{i \Omega k t}\Tr \mu \hat \rho_n^k\\
  &= \frac{1}{2\pi}\sum_{k,k'=0}^{\infty}  E^*_{\eta,k'} \Tr \mu \hat \rho_n^k e^{i \Omega (k - k') t}.
\end{align}
The integrated heterodyne signal can then be approximated as above by
a finite time integral,
\begin{align}
  I^{(n)}_\text{T, het} = \frac{1}{T} \int_{0}^{T}  \mathrm dt I^{(n)}_\text{het}(t) = \frac{1}{2\pi}\sum_{k=0}^{\infty}  E^*_{\eta,k} \Tr \mu \hat \rho_n^k. \label{eq:heterodyne-integ}
\end{align}
This is the signal as measured in Fig.~\ref{fig:pyrlinear} or in a
pump-probe experiment.  The heterodyne signal can also be dispersed
through a monochromator to obtain a frequency-resolved measurement, as
is done in transient absorption spectroscopy.  The monochromated
quantities at output frequency $\Omega k_\text{out}$ are given by,
\begin{align}
\mu_{T,n}(t, k_\text{out}) &= \frac{e^{\eta t}}{2\pi} e^{i \Omega k_\text{out} t}\Tr \mu \hat \rho_n^{k_\text{out}}\\
 E^*_+(t, k_\text{out}) &= E^*_{k_\text{out}} e^{-i \Omega  k_\text{out} t} 
\end{align}
where $k_\text{out}$ is the frequency of the monochromator.  Then, the
integrated signal is given by
\begin{align}
  I^{(n)}_\text{T, het}(k_\text{out}) &= \frac{1}{T} \int_{0}^{T} E^*_+(t, k_\text{out}) \mu_{T,n}(t, k_\text{out}) \\
  &=\frac{1}{2\pi} E^*_{\eta, k_\text{out}} \Tr \mu \hat \rho_n^{k_\text{out}}\label{eq:heterodyne}
\end{align}
This formula is used to obtain the transient absorption spectra above.

\end{document}

%% file: lambda.tex
\begin{tikzpicture}[
  scale=0.6,
  decoration={snake, pre length=2pt, post length=5pt, amplitude=0.5mm},
  level/.style={thick},
  virtual/.style={thick,densely dashed},
  trans/.style={double,<->},
  relax/.style={->,decorate,>=stealth'},
  ]

  \draw[level] (-2,0) -- (0,0) node[midway, below] {$\ket{g}$};
  \draw[level] (2,1) -- (4,1) node[midway, below] {$\ket{g'}$};
  \draw[level] (-1,3.75) -- (1,3.75);
  \path (-0.2, 4)  node[inner sep=0](mid1){} --
  (0.2, 4) node[inner sep=0](mid2){}; 
  \draw[level] (-1,4.25) -- (1,4.25) node[midway, above] {$\ket{e_1},\ket{e_2}$};

  \draw[trans] (-1,0) -- (mid1) node [midway,left]{$\mu_{g,e_i}$};
  \draw[trans] (3,1) -- (mid2) node [midway,right]{$\mu_{g',e_i}$};

\end{tikzpicture}

%% file: pump-probe.tex
\begin{tikzpicture}[
  domain=-6:2,
  scale=0.8,
  decoration={snake, pre length=2pt, post length=5pt, amplitude=0.5mm},
  path/.style={very thick, ->, color=gray, >=stealth'},
  photon/.style={->,decorate,>=stealth'},
  excite/.style={ ->, thick}]
  \def\w{0.0739};
  \def\Eone{3.94};
  \def\Etwo{4.89};
  \def\kone{-0.09806};
  \def\ktwo{ 0.13545};
  \def\shift{0.1};
  \def\xci{-4.06};
  \def\yci{4.950}

  \draw[thick] (-6,0) -- (2,0) node[below, midway]{$\ket{g}$};

  \draw[variable=\x,smooth] plot[samples=100] ({\x}, {0.5 * \w * \x * \x  + \kone * \x + \Eone}) node[right] {$S_1$}; 

  \draw[variable=\x,smooth] plot[samples=100] ({\x}, {0.5 * \w * \x * \x  + \ktwo * \x + \Etwo}) node[right] {$S_2$}; 
  \draw[path,variable=\x,smooth,domain=0:\xci+0.2] plot[samples=100] ({\x}, {0.5 * \w * \x * \x  + \ktwo * \x + \Etwo + \shift}); 
  \draw[path,densely dashed,variable=\x,smooth,domain=\xci+0.2:2] plot[samples=100] ({\x}, {0.5 * \w * \x * \x  + \ktwo * \x + \Etwo - \shift}); 
  \draw[path,densely dashed, variable=\x,smooth,domain=\xci-0.2:-6] plot[samples=100] ({\x}, {0.5 * \w * \x * \x  + \ktwo * \x + \Etwo -  \shift}); 

  \node[above] at (-2, 4.8) {\textsf{A}};
  \node[below] at (-5, 5) {\textsf{B}};
  \node[below] at (1, 5) {\textsf{C}};

  \draw[excite] (0,0) -- (0, \Etwo) node[above] {FC};
  \draw[photon] (2,2)  -- (0, 2) node[midway,below] {Pump};

  \draw[excite] (-2,0) -- (-2, 4.73);
  \draw[photon] (-4,2)  -- (-2, 2)node[midway,below] {Probe};

  \node[above] at (\xci, \yci) {CI};
\end{tikzpicture}



%% file: paper.bbl
\begin{thebibliography}{66}%
\makeatletter
\providecommand \@ifxundefined [1]{%
 \@ifx{#1\undefined}
}%
\providecommand \@ifnum [1]{%
 \ifnum #1\expandafter \@firstoftwo
 \else \expandafter \@secondoftwo
 \fi
}%
\providecommand \@ifx [1]{%
 \ifx #1\expandafter \@firstoftwo
 \else \expandafter \@secondoftwo
 \fi
}%
\providecommand \natexlab [1]{#1}%
\providecommand \enquote  [1]{``#1''}%
\providecommand \bibnamefont  [1]{#1}%
\providecommand \bibfnamefont [1]{#1}%
\providecommand \citenamefont [1]{#1}%
\providecommand \href@noop [0]{\@secondoftwo}%
\providecommand \href [0]{\begingroup \@sanitize@url \@href}%
\providecommand \@href[1]{\@@startlink{#1}\@@href}%
\providecommand \@@href[1]{\endgroup#1\@@endlink}%
\providecommand \@sanitize@url [0]{\catcode `\\12\catcode `\$12\catcode
  `\&12\catcode `\#12\catcode `\^12\catcode `\_12\catcode `\%12\relax}%
\providecommand \@@startlink[1]{}%
\providecommand \@@endlink[0]{}%
\providecommand \url  [0]{\begingroup\@sanitize@url \@url }%
\providecommand \@url [1]{\endgroup\@href {#1}{\urlprefix }}%
\providecommand \urlprefix  [0]{URL }%
\providecommand \Eprint [0]{\href }%
\providecommand \doibase [0]{http://dx.doi.org/}%
\providecommand \selectlanguage [0]{\@gobble}%
\providecommand \bibinfo  [0]{\@secondoftwo}%
\providecommand \bibfield  [0]{\@secondoftwo}%
\providecommand \translation [1]{[#1]}%
\providecommand \BibitemOpen [0]{}%
\providecommand \bibitemStop [0]{}%
\providecommand \bibitemNoStop [0]{.\EOS\space}%
\providecommand \EOS [0]{\spacefactor3000\relax}%
\providecommand \BibitemShut  [1]{\csname bibitem#1\endcsname}%
\let\auto@bib@innerbib\@empty
\bibitem [{\citenamefont {Shapiro}\ and\ \citenamefont
  {Brumer}(2012)}]{shapiro_quantum_2012}%
  \BibitemOpen
  \bibfield  {author} {\bibinfo {author} {\bibfnamefont {M.}~\bibnamefont
  {Shapiro}}\ and\ \bibinfo {author} {\bibfnamefont {P.}~\bibnamefont
  {Brumer}},\ }\href@noop {} {\emph {\bibinfo {title} {Quantum {{Control}} of
  {{Molecular Processes}}}}}\ (\bibinfo  {publisher} {{John Wiley \& Sons}},\
  \bibinfo {year} {2012})\BibitemShut {NoStop}%
\bibitem [{\citenamefont {{Cohen-Tannoudji}}, \citenamefont {{Dupont-Roc}},\
  and\ \citenamefont {Grynberg}(1992)}]{cohen-tannoudji_atom-photon_1992}%
  \BibitemOpen
  \bibfield  {author} {\bibinfo {author} {\bibfnamefont {C.}~\bibnamefont
  {{Cohen-Tannoudji}}}, \bibinfo {author} {\bibfnamefont {J.}~\bibnamefont
  {{Dupont-Roc}}}, \ and\ \bibinfo {author} {\bibfnamefont {G.}~\bibnamefont
  {Grynberg}},\ }\href@noop {} {\emph {\bibinfo {title} {Atom-Photon
  Interactions: Basic Processes and Applications}}}\ (\bibinfo  {publisher}
  {{J. Wiley}},\ \bibinfo {year} {1992})\BibitemShut {NoStop}%
\bibitem [{\citenamefont {Gallagher~Faeder}\ and\ \citenamefont
  {Jonas}(1999)}]{gallagher_faeder_two-dimensional_1999}%
  \BibitemOpen
  \bibfield  {author} {\bibinfo {author} {\bibfnamefont {S.~M.}\ \bibnamefont
  {Gallagher~Faeder}}\ and\ \bibinfo {author} {\bibfnamefont {D.~M.}\
  \bibnamefont {Jonas}},\ }\href {\doibase 10.1021/jp9925738} {\bibfield
  {journal} {\bibinfo  {journal} {J. Phys. Chem. A}\ }\textbf {\bibinfo
  {volume} {103}},\ \bibinfo {pages} {10489} (\bibinfo {year}
  {1999})}\BibitemShut {NoStop}%
\bibitem [{\citenamefont {Zhuang}\ \emph {et~al.}(2006)\citenamefont {Zhuang},
  \citenamefont {Abramavicius}, \citenamefont {Hayashi},\ and\ \citenamefont
  {Mukamel}}]{zhuang_simulation_2006}%
  \BibitemOpen
  \bibfield  {author} {\bibinfo {author} {\bibfnamefont {W.}~\bibnamefont
  {Zhuang}}, \bibinfo {author} {\bibfnamefont {D.}~\bibnamefont
  {Abramavicius}}, \bibinfo {author} {\bibfnamefont {T.}~\bibnamefont
  {Hayashi}}, \ and\ \bibinfo {author} {\bibfnamefont {S.}~\bibnamefont
  {Mukamel}},\ }\href {\doibase 10.1021/jp055813u} {\bibfield  {journal}
  {\bibinfo  {journal} {J. Phys. Chem. B}\ }\textbf {\bibinfo {volume} {110}},\
  \bibinfo {pages} {3362} (\bibinfo {year} {2006})}\BibitemShut {NoStop}%
\bibitem [{\citenamefont {Quesada}\ and\ \citenamefont
  {Sipe}(2014)}]{quesada_effects_2014}%
  \BibitemOpen
  \bibfield  {author} {\bibinfo {author} {\bibfnamefont {N.}~\bibnamefont
  {Quesada}}\ and\ \bibinfo {author} {\bibfnamefont {J.~E.}\ \bibnamefont
  {Sipe}},\ }\href {\doibase 10.1103/PhysRevA.90.063840} {\bibfield  {journal}
  {\bibinfo  {journal} {Phys. Rev. A}\ }\textbf {\bibinfo {volume} {90}},\
  \bibinfo {pages} {063840} (\bibinfo {year} {2014})}\BibitemShut {NoStop}%
\bibitem [{\citenamefont {Br{\"u}hl}, \citenamefont {Buckup},\ and\
  \citenamefont {Motzkus}(2018)}]{bruhl_experimental_2018}%
  \BibitemOpen
  \bibfield  {author} {\bibinfo {author} {\bibfnamefont {E.}~\bibnamefont
  {Br{\"u}hl}}, \bibinfo {author} {\bibfnamefont {T.}~\bibnamefont {Buckup}}, \
  and\ \bibinfo {author} {\bibfnamefont {M.}~\bibnamefont {Motzkus}},\ }\href
  {\doibase 10.1063/1.5029805} {\bibfield  {journal} {\bibinfo  {journal} {J.
  Chem. Phys.}\ }\textbf {\bibinfo {volume} {148}},\ \bibinfo {pages} {214310}
  (\bibinfo {year} {2018})}\BibitemShut {NoStop}%
\bibitem [{\citenamefont {Reppert}\ and\ \citenamefont
  {Brumer}(2018{\natexlab{a}})}]{reppert_classical_2018}%
  \BibitemOpen
  \bibfield  {author} {\bibinfo {author} {\bibfnamefont {M.}~\bibnamefont
  {Reppert}}\ and\ \bibinfo {author} {\bibfnamefont {P.}~\bibnamefont
  {Brumer}},\ }\href {\doibase 10.1063/1.5017985} {\bibfield  {journal}
  {\bibinfo  {journal} {J. Chem. Phys.}\ }\textbf {\bibinfo {volume} {148}},\
  \bibinfo {pages} {064101} (\bibinfo {year} {2018}{\natexlab{a}})}\BibitemShut
  {NoStop}%
\bibitem [{\citenamefont {Pach{\'o}n}\ and\ \citenamefont
  {Brumer}(2013)}]{pachon_mechanisms_2013}%
  \BibitemOpen
  \bibfield  {author} {\bibinfo {author} {\bibfnamefont {L.~A.}\ \bibnamefont
  {Pach{\'o}n}}\ and\ \bibinfo {author} {\bibfnamefont {P.}~\bibnamefont
  {Brumer}},\ }\href {\doibase 10.1063/1.4825358} {\bibfield  {journal}
  {\bibinfo  {journal} {J. Chem. Phys.}\ }\textbf {\bibinfo {volume} {139}},\
  \bibinfo {pages} {164123} (\bibinfo {year} {2013})}\BibitemShut {NoStop}%
\bibitem [{\citenamefont {Mukamel}(2013)}]{mukamel_coherent-control_2013}%
  \BibitemOpen
  \bibfield  {author} {\bibinfo {author} {\bibfnamefont {S.}~\bibnamefont
  {Mukamel}},\ }\href {\doibase 10.1063/1.4824857} {\bibfield  {journal}
  {\bibinfo  {journal} {J. Chem. Phys.}\ }\textbf {\bibinfo {volume} {139}},\
  \bibinfo {pages} {164113} (\bibinfo {year} {2013})}\BibitemShut {NoStop}%
\bibitem [{\citenamefont {{Am-Shallem}}\ and\ \citenamefont
  {Kosloff}(2014)}]{am-shallem_scaling_2014}%
  \BibitemOpen
  \bibfield  {author} {\bibinfo {author} {\bibfnamefont {M.}~\bibnamefont
  {{Am-Shallem}}}\ and\ \bibinfo {author} {\bibfnamefont {R.}~\bibnamefont
  {Kosloff}},\ }\href {\doibase 10.1063/1.4890822} {\bibfield  {journal}
  {\bibinfo  {journal} {J. Chem. Phys.}\ }\textbf {\bibinfo {volume} {141}},\
  \bibinfo {pages} {044121} (\bibinfo {year} {2014})}\BibitemShut {NoStop}%
\bibitem [{\citenamefont {Lavigne}\ and\ \citenamefont
  {Brumer}(2017)}]{lavigne_interfering_2017}%
  \BibitemOpen
  \bibfield  {author} {\bibinfo {author} {\bibfnamefont {C.}~\bibnamefont
  {Lavigne}}\ and\ \bibinfo {author} {\bibfnamefont {P.}~\bibnamefont
  {Brumer}},\ }\href {\doibase 10.1063/1.5003389} {\bibfield  {journal}
  {\bibinfo  {journal} {J. Chem. Phys.}\ }\textbf {\bibinfo {volume} {147}},\
  \bibinfo {pages} {114107} (\bibinfo {year} {2017})}\BibitemShut {NoStop}%
\bibitem [{\citenamefont {Lavigne}\ and\ \citenamefont
  {Brumer}(2019{\natexlab{a}})}]{lavigne_two-photon_2019}%
  \BibitemOpen
  \bibfield  {author} {\bibinfo {author} {\bibfnamefont {C.}~\bibnamefont
  {Lavigne}}\ and\ \bibinfo {author} {\bibfnamefont {P.}~\bibnamefont
  {Brumer}},\ }\href@noop {} {\bibfield  {journal} {\bibinfo  {journal} {In
  prep.}\ } (\bibinfo {year} {2019}{\natexlab{a}})}\BibitemShut {NoStop}%
\bibitem [{\citenamefont {Tscherbul}\ and\ \citenamefont
  {Brumer}(2014)}]{tscherbul_excitation_2014}%
  \BibitemOpen
  \bibfield  {author} {\bibinfo {author} {\bibfnamefont {T.~V.}\ \bibnamefont
  {Tscherbul}}\ and\ \bibinfo {author} {\bibfnamefont {P.}~\bibnamefont
  {Brumer}},\ }\href {\doibase 10.1021/jp501700t} {\bibfield  {journal}
  {\bibinfo  {journal} {J. Phys. Chem. A}\ }\textbf {\bibinfo {volume} {118}},\
  \bibinfo {pages} {3100} (\bibinfo {year} {2014})}\BibitemShut {NoStop}%
\bibitem [{\citenamefont {Tscherbul}\ and\ \citenamefont
  {Brumer}(2015)}]{tscherbul_quantum_2015}%
  \BibitemOpen
  \bibfield  {author} {\bibinfo {author} {\bibfnamefont {T.~V.}\ \bibnamefont
  {Tscherbul}}\ and\ \bibinfo {author} {\bibfnamefont {P.}~\bibnamefont
  {Brumer}},\ }\href {\doibase 10.1039/C5CP01388G} {\bibfield  {journal}
  {\bibinfo  {journal} {Phys. Chem. Chem. Phys.}\ }\textbf {\bibinfo {volume}
  {17}},\ \bibinfo {pages} {30904} (\bibinfo {year} {2015})}\BibitemShut
  {NoStop}%
\bibitem [{\citenamefont {Brumer}(2018)}]{brumer_shedding_2018}%
  \BibitemOpen
  \bibfield  {author} {\bibinfo {author} {\bibfnamefont {P.}~\bibnamefont
  {Brumer}},\ }\href {\doibase 10.1021/acs.jpclett.8b00874} {\bibfield
  {journal} {\bibinfo  {journal} {J. Phys. Chem. Lett.}\ }\textbf {\bibinfo
  {volume} {9}},\ \bibinfo {pages} {2946} (\bibinfo {year} {2018})}\BibitemShut
  {NoStop}%
\bibitem [{\citenamefont {Reppert}\ and\ \citenamefont
  {Brumer}(2018{\natexlab{b}})}]{reppert_quantumness_2018}%
  \BibitemOpen
  \bibfield  {author} {\bibinfo {author} {\bibfnamefont {M.}~\bibnamefont
  {Reppert}}\ and\ \bibinfo {author} {\bibfnamefont {P.}~\bibnamefont
  {Brumer}},\ }\href {\doibase 10.1063/1.5058136} {\bibfield  {journal}
  {\bibinfo  {journal} {J. Chem. Phys.}\ }\textbf {\bibinfo {volume} {149}},\
  \bibinfo {pages} {234102} (\bibinfo {year} {2018}{\natexlab{b}})}\BibitemShut
  {NoStop}%
\bibitem [{\citenamefont {Smallwood}, \citenamefont {Autry},\ and\
  \citenamefont {Cundiff}(2017)}]{smallwood_analytical_2017}%
  \BibitemOpen
  \bibfield  {author} {\bibinfo {author} {\bibfnamefont {C.~L.}\ \bibnamefont
  {Smallwood}}, \bibinfo {author} {\bibfnamefont {T.~M.}\ \bibnamefont
  {Autry}}, \ and\ \bibinfo {author} {\bibfnamefont {S.~T.}\ \bibnamefont
  {Cundiff}},\ }\href {\doibase 10.1364/JOSAB.34.000419} {\bibfield  {journal}
  {\bibinfo  {journal} {J. Opt. Soc. Am. B, JOSAB}\ }\textbf {\bibinfo {volume}
  {34}},\ \bibinfo {pages} {419} (\bibinfo {year} {2017})}\BibitemShut
  {NoStop}%
\bibitem [{\citenamefont {Perl{\'i}k}, \citenamefont {Hauer},\ and\
  \citenamefont {{\v S}anda}(2017)}]{perlik_finite_2017}%
  \BibitemOpen
  \bibfield  {author} {\bibinfo {author} {\bibfnamefont {V.}~\bibnamefont
  {Perl{\'i}k}}, \bibinfo {author} {\bibfnamefont {J.}~\bibnamefont {Hauer}}, \
  and\ \bibinfo {author} {\bibfnamefont {F.}~\bibnamefont {{\v S}anda}},\
  }\href {\doibase 10.1364/JOSAB.34.000430} {\bibfield  {journal} {\bibinfo
  {journal} {J. Opt. Soc. Am. B, JOSAB}\ }\textbf {\bibinfo {volume} {34}},\
  \bibinfo {pages} {430} (\bibinfo {year} {2017})}\BibitemShut {NoStop}%
\bibitem [{\citenamefont {Brumer}\ and\ \citenamefont
  {Shapiro}(1992)}]{brumer_laser_1992}%
  \BibitemOpen
  \bibfield  {author} {\bibinfo {author} {\bibfnamefont {P.}~\bibnamefont
  {Brumer}}\ and\ \bibinfo {author} {\bibfnamefont {M.}~\bibnamefont
  {Shapiro}},\ }\href {\doibase 10.1146/annurev.pc.43.100192.001353} {\bibfield
   {journal} {\bibinfo  {journal} {Annu. Rev. Phys. Chem.}\ }\textbf {\bibinfo
  {volume} {43}},\ \bibinfo {pages} {257} (\bibinfo {year} {1992})}\BibitemShut
  {NoStop}%
\bibitem [{\citenamefont {Spanner}, \citenamefont {Arango},\ and\ \citenamefont
  {Brumer}(2010)}]{spanner_communication_2010}%
  \BibitemOpen
  \bibfield  {author} {\bibinfo {author} {\bibfnamefont {M.}~\bibnamefont
  {Spanner}}, \bibinfo {author} {\bibfnamefont {C.~A.}\ \bibnamefont {Arango}},
  \ and\ \bibinfo {author} {\bibfnamefont {P.}~\bibnamefont {Brumer}},\ }\href
  {\doibase 10.1063/1.3491366} {\bibfield  {journal} {\bibinfo  {journal} {J.
  Chem. Phys.}\ }\textbf {\bibinfo {volume} {133}},\ \bibinfo {pages} {151101}
  (\bibinfo {year} {2010})}\BibitemShut {NoStop}%
\bibitem [{\citenamefont {Faisal}(1987)}]{faisal_theory_1987}%
  \BibitemOpen
  \bibfield  {author} {\bibinfo {author} {\bibfnamefont {F.~H.~M.}\
  \bibnamefont {Faisal}},\ }\href@noop {} {\emph {\bibinfo {title} {Theory of
  {{Multiphoton Processes}}}}}\ (\bibinfo  {publisher} {{Springer US}},\
  \bibinfo {year} {1987})\BibitemShut {NoStop}%
\bibitem [{\citenamefont {Mukamel}(1995)}]{mukamel_principles_1995}%
  \BibitemOpen
  \bibfield  {author} {\bibinfo {author} {\bibfnamefont {S.}~\bibnamefont
  {Mukamel}},\ }\href@noop {} {\emph {\bibinfo {title} {Principles of Nonlinear
  Optical Spectroscopy}}}\ (\bibinfo  {publisher} {{Oxford University Press}},\
  \bibinfo {year} {1995})\BibitemShut {NoStop}%
\bibitem [{\citenamefont {Plenio}, \citenamefont {Almeida},\ and\ \citenamefont
  {Huelga}(2013)}]{plenio_origin_2013}%
  \BibitemOpen
  \bibfield  {author} {\bibinfo {author} {\bibfnamefont {M.~B.}\ \bibnamefont
  {Plenio}}, \bibinfo {author} {\bibfnamefont {J.}~\bibnamefont {Almeida}}, \
  and\ \bibinfo {author} {\bibfnamefont {S.~F.}\ \bibnamefont {Huelga}},\
  }\href {\doibase 10.1063/1.4846275} {\bibfield  {journal} {\bibinfo
  {journal} {J. Chem. Phys.}\ }\textbf {\bibinfo {volume} {139}},\ \bibinfo
  {pages} {235102} (\bibinfo {year} {2013})}\BibitemShut {NoStop}%
\bibitem [{\citenamefont {Katz}, \citenamefont {Ratner},\ and\ \citenamefont
  {Kosloff}(2010)}]{katz_control_2010}%
  \BibitemOpen
  \bibfield  {author} {\bibinfo {author} {\bibfnamefont {G.}~\bibnamefont
  {Katz}}, \bibinfo {author} {\bibfnamefont {M.~A.}\ \bibnamefont {Ratner}}, \
  and\ \bibinfo {author} {\bibfnamefont {R.}~\bibnamefont {Kosloff}},\ }\href
  {\doibase 10.1088/1367-2630/12/1/015003} {\bibfield  {journal} {\bibinfo
  {journal} {New J. Phys.}\ }\textbf {\bibinfo {volume} {12}},\ \bibinfo
  {pages} {015003} (\bibinfo {year} {2010})}\BibitemShut {NoStop}%
\bibitem [{\citenamefont {Arango}\ and\ \citenamefont
  {Brumer}(2013)}]{arango_communication_2013}%
  \BibitemOpen
  \bibfield  {author} {\bibinfo {author} {\bibfnamefont {C.~A.}\ \bibnamefont
  {Arango}}\ and\ \bibinfo {author} {\bibfnamefont {P.}~\bibnamefont
  {Brumer}},\ }\href {\doibase 10.1063/1.4792834} {\bibfield  {journal}
  {\bibinfo  {journal} {J. Chem. Phys.}\ }\textbf {\bibinfo {volume} {138}},\
  \bibinfo {pages} {071104} (\bibinfo {year} {2013})}\BibitemShut {NoStop}%
\bibitem [{\citenamefont {Abe}\ \emph {et~al.}(2005)\citenamefont {Abe},
  \citenamefont {Ohtsuki}, \citenamefont {Fujimura},\ and\ \citenamefont
  {Domcke}}]{abe_optimal_2005}%
  \BibitemOpen
  \bibfield  {author} {\bibinfo {author} {\bibfnamefont {M.}~\bibnamefont
  {Abe}}, \bibinfo {author} {\bibfnamefont {Y.}~\bibnamefont {Ohtsuki}},
  \bibinfo {author} {\bibfnamefont {Y.}~\bibnamefont {Fujimura}}, \ and\
  \bibinfo {author} {\bibfnamefont {W.}~\bibnamefont {Domcke}},\ }\href
  {\doibase 10.1063/1.2034488} {\bibfield  {journal} {\bibinfo  {journal} {J.
  Chem. Phys.}\ }\textbf {\bibinfo {volume} {123}},\ \bibinfo {pages} {144508}
  (\bibinfo {year} {2005})}\BibitemShut {NoStop}%
\bibitem [{\citenamefont {Milonni}\ and\ \citenamefont
  {Eberly}(2010)}]{milonni_laser_2010}%
  \BibitemOpen
  \bibfield  {author} {\bibinfo {author} {\bibfnamefont {P.~W.}\ \bibnamefont
  {Milonni}}\ and\ \bibinfo {author} {\bibfnamefont {J.~H.}\ \bibnamefont
  {Eberly}},\ }\href@noop {} {\emph {\bibinfo {title} {Laser {{Physics}}}}}\
  (\bibinfo  {publisher} {{John Wiley \& Sons}},\ \bibinfo {year}
  {2010})\BibitemShut {NoStop}%
\bibitem [{\citenamefont {Han}\ and\ \citenamefont
  {Shapiro}(2012)}]{han_linear_2012}%
  \BibitemOpen
  \bibfield  {author} {\bibinfo {author} {\bibfnamefont {A.~C.}\ \bibnamefont
  {Han}}\ and\ \bibinfo {author} {\bibfnamefont {M.}~\bibnamefont {Shapiro}},\
  }\href {\doibase 10.1103/PhysRevLett.108.183002} {\bibfield  {journal}
  {\bibinfo  {journal} {Phys. Rev. Lett.}\ }\textbf {\bibinfo {volume} {108}},\
  \bibinfo {pages} {183002} (\bibinfo {year} {2012})}\BibitemShut {NoStop}%
\bibitem [{\citenamefont {Han}\ and\ \citenamefont
  {Shapiro}(2013)}]{han_linear_2013}%
  \BibitemOpen
  \bibfield  {author} {\bibinfo {author} {\bibfnamefont {A.~C.}\ \bibnamefont
  {Han}}\ and\ \bibinfo {author} {\bibfnamefont {M.}~\bibnamefont {Shapiro}},\
  }\href {\doibase 10.1088/0953-4075/46/8/085401} {\bibfield  {journal}
  {\bibinfo  {journal} {J. Phys. B: At. Mol. Opt. Phys.}\ }\textbf {\bibinfo
  {volume} {46}},\ \bibinfo {pages} {085401} (\bibinfo {year}
  {2013})}\BibitemShut {NoStop}%
\bibitem [{\citenamefont {Lavigne}\ and\ \citenamefont
  {Brumer}(2019{\natexlab{b}})}]{lavigne_ultrafast_2019}%
  \BibitemOpen
  \bibfield  {author} {\bibinfo {author} {\bibfnamefont {C.}~\bibnamefont
  {Lavigne}}\ and\ \bibinfo {author} {\bibfnamefont {P.}~\bibnamefont
  {Brumer}},\ }\href@noop {} {\bibfield  {journal} {\bibinfo  {journal} {In
  prep.}\ } (\bibinfo {year} {2019}{\natexlab{b}})}\BibitemShut {NoStop}%
\bibitem [{\citenamefont {Meyer}\ and\ \citenamefont
  {Engel}(2000)}]{meyer_non-perturbative_2000}%
  \BibitemOpen
  \bibfield  {author} {\bibinfo {author} {\bibfnamefont {S.}~\bibnamefont
  {Meyer}}\ and\ \bibinfo {author} {\bibfnamefont {V.}~\bibnamefont {Engel}},\
  }\href {\doibase 10.1007/s003400000342} {\bibfield  {journal} {\bibinfo
  {journal} {Appl Phys B}\ }\textbf {\bibinfo {volume} {71}},\ \bibinfo {pages}
  {293} (\bibinfo {year} {2000})}\BibitemShut {NoStop}%
\bibitem [{\citenamefont {Dubner}\ and\ \citenamefont
  {Abate}(1968)}]{dubner_numerical_1968}%
  \BibitemOpen
  \bibfield  {author} {\bibinfo {author} {\bibfnamefont {H.}~\bibnamefont
  {Dubner}}\ and\ \bibinfo {author} {\bibfnamefont {J.}~\bibnamefont {Abate}},\
  }\href {\doibase 10.1145/321439.321446} {\bibfield  {journal} {\bibinfo
  {journal} {J. ACM}\ }\textbf {\bibinfo {volume} {15}},\ \bibinfo {pages}
  {115} (\bibinfo {year} {1968})}\BibitemShut {NoStop}%
\bibitem [{\citenamefont {Veillon}(1974)}]{veillon_algorithm_1974}%
  \BibitemOpen
  \bibfield  {author} {\bibinfo {author} {\bibfnamefont {F.}~\bibnamefont
  {Veillon}},\ }\href {\doibase 10.1145/355620.361174} {\bibfield  {journal}
  {\bibinfo  {journal} {Commun. ACM}\ }\textbf {\bibinfo {volume} {17}},\
  \bibinfo {pages} {587} (\bibinfo {year} {1974})}\BibitemShut {NoStop}%
\bibitem [{\citenamefont {Crump}(1976)}]{crump_numerical_1976}%
  \BibitemOpen
  \bibfield  {author} {\bibinfo {author} {\bibfnamefont {K.~S.}\ \bibnamefont
  {Crump}},\ }\href {\doibase 10.1145/321921.321931} {\bibfield  {journal}
  {\bibinfo  {journal} {J. ACM}\ }\textbf {\bibinfo {volume} {23}},\ \bibinfo
  {pages} {89} (\bibinfo {year} {1976})}\BibitemShut {NoStop}%
\bibitem [{\citenamefont {{de Hoog}}, \citenamefont {Knight},\ and\
  \citenamefont {Stokes}(1982)}]{de_hoog_improved_1982}%
  \BibitemOpen
  \bibfield  {author} {\bibinfo {author} {\bibfnamefont {F.~R.}\ \bibnamefont
  {{de Hoog}}}, \bibinfo {author} {\bibfnamefont {J.}~\bibnamefont {Knight}}, \
  and\ \bibinfo {author} {\bibfnamefont {A.}~\bibnamefont {Stokes}},\ }\href
  {\doibase 10.1137/0903022} {\bibfield  {journal} {\bibinfo  {journal} {SIAM
  J. Sci. Stat. Comp.}\ }\textbf {\bibinfo {volume} {3}},\ \bibinfo {pages}
  {357} (\bibinfo {year} {1982})}\BibitemShut {NoStop}%
\bibitem [{\citenamefont {Piessens}\ and\ \citenamefont
  {Huysmans}(1984)}]{piessens_algorithm_1984}%
  \BibitemOpen
  \bibfield  {author} {\bibinfo {author} {\bibfnamefont {R.}~\bibnamefont
  {Piessens}}\ and\ \bibinfo {author} {\bibfnamefont {R.}~\bibnamefont
  {Huysmans}},\ }\href {\doibase 10.1145/1271.319416} {\bibfield  {journal}
  {\bibinfo  {journal} {ACM T. Math. Software}\ }\textbf {\bibinfo {volume}
  {10}},\ \bibinfo {pages} {348} (\bibinfo {year} {1984})}\BibitemShut
  {NoStop}%
\bibitem [{\citenamefont {Gerstner}\ and\ \citenamefont
  {Griebel}(1998)}]{gerstner_numerical_1998}%
  \BibitemOpen
  \bibfield  {author} {\bibinfo {author} {\bibfnamefont {T.}~\bibnamefont
  {Gerstner}}\ and\ \bibinfo {author} {\bibfnamefont {M.}~\bibnamefont
  {Griebel}},\ }\href {\doibase 10.1023/A:1019129717644} {\bibfield  {journal}
  {\bibinfo  {journal} {Numerical Algorithms}\ }\textbf {\bibinfo {volume}
  {18}},\ \bibinfo {pages} {209} (\bibinfo {year} {1998})}\BibitemShut
  {NoStop}%
\bibitem [{\citenamefont {Lavigne}(2019)}]{lavigne_flipt.jl_2019}%
  \BibitemOpen
  \bibfield  {author} {\bibinfo {author} {\bibfnamefont {C.}~\bibnamefont
  {Lavigne}},\ }\href@noop {} {\enquote {\bibinfo {title} {{{FLIPT}}.jl, a
  highly efficient method to compute the {{Fourier}}-{{Laplace Inversion}} of
  the {{Perturbation Theory}}},}\ } (\bibinfo {year} {2019})\BibitemShut
  {NoStop}%
\bibitem [{\citenamefont {Axelrod}\ and\ \citenamefont
  {Brumer}(2018)}]{axelrod_efficient_2018}%
  \BibitemOpen
  \bibfield  {author} {\bibinfo {author} {\bibfnamefont {S.}~\bibnamefont
  {Axelrod}}\ and\ \bibinfo {author} {\bibfnamefont {P.}~\bibnamefont
  {Brumer}},\ }\href {\doibase 10.1063/1.5041005} {\bibfield  {journal}
  {\bibinfo  {journal} {J. Chem. Phys.}\ }\textbf {\bibinfo {volume} {149}},\
  \bibinfo {pages} {114104} (\bibinfo {year} {2018})}\BibitemShut {NoStop}%
\bibitem [{Note1()}]{Note1}%
  \BibitemOpen
  \bibinfo {note} {The algorithm can easily be extended to the case where the
  perturbation is composed of multiple components $\DOTSB \sum@ \slimits@
  _\alpha E_\alpha (t) V_\alpha (t)$ by computing and summing over all unique
  combinations of $\alpha $.}\BibitemShut {Stop}%
\bibitem [{\citenamefont {Lendi}(1977)}]{lendi_superoperator_1977}%
  \BibitemOpen
  \bibfield  {author} {\bibinfo {author} {\bibfnamefont {K.}~\bibnamefont
  {Lendi}},\ }\href {\doibase 10.1016/0301-0104(77)85121-5} {\bibfield
  {journal} {\bibinfo  {journal} {Chemical Physics}\ }\textbf {\bibinfo
  {volume} {20}},\ \bibinfo {pages} {135} (\bibinfo {year} {1977})}\BibitemShut
  {NoStop}%
\bibitem [{\citenamefont {Boas}(2005)}]{boas_mathematical_2005}%
  \BibitemOpen
  \bibfield  {author} {\bibinfo {author} {\bibfnamefont {M.~L.}\ \bibnamefont
  {Boas}},\ }\href@noop {} {\emph {\bibinfo {title} {Mathematical {{Methods}}
  in the {{Physical Sciences}}}}}\ (\bibinfo  {publisher} {{Wiley}},\ \bibinfo
  {year} {2005})\BibitemShut {NoStop}%
\bibitem [{\citenamefont {L{\"o}wdin}(1982)}]{lowdin_operators_1982}%
  \BibitemOpen
  \bibfield  {author} {\bibinfo {author} {\bibfnamefont {P.-O.}\ \bibnamefont
  {L{\"o}wdin}},\ }\href {\doibase 10.1002/qua.560220847} {\bibfield  {journal}
  {\bibinfo  {journal} {International Journal of Quantum Chemistry}\ }\textbf
  {\bibinfo {volume} {22}},\ \bibinfo {pages} {485} (\bibinfo {year}
  {1982})}\BibitemShut {NoStop}%
\bibitem [{Note2()}]{Note2}%
  \BibitemOpen
  \bibinfo {note} {The case where the initial state is not a steady state can
  be computed directly from eq.~(\ref {eq:rec1}) and (\ref {eq:rec2}) above.
  This is numerically more expensive as it lead to an $n+1$ dimensional
  integral for the $n$-th perturbative term instead of an $n$ dimensional
  integral.}\BibitemShut {Stop}%
\bibitem [{\citenamefont {Rose}\ and\ \citenamefont
  {Krich}(2019)}]{rose_numerical_2019}%
  \BibitemOpen
  \bibfield  {author} {\bibinfo {author} {\bibfnamefont {P.~A.}\ \bibnamefont
  {Rose}}\ and\ \bibinfo {author} {\bibfnamefont {J.~J.}\ \bibnamefont
  {Krich}},\ }\href@noop {} {\bibfield  {journal} {\bibinfo  {journal}
  {arXiv:1902.07854 [physics]}\ } (\bibinfo {year} {2019})},\ \Eprint
  {http://arxiv.org/abs/1902.07854} {arXiv:1902.07854 [physics]} \BibitemShut
  {NoStop}%
\bibitem [{\citenamefont {{de Vega}}\ and\ \citenamefont
  {Alonso}(2017)}]{de_vega_dynamics_2017}%
  \BibitemOpen
  \bibfield  {author} {\bibinfo {author} {\bibfnamefont {I.}~\bibnamefont {{de
  Vega}}}\ and\ \bibinfo {author} {\bibfnamefont {D.}~\bibnamefont {Alonso}},\
  }\href {\doibase 10.1103/RevModPhys.89.015001} {\bibfield  {journal}
  {\bibinfo  {journal} {Rev. Mod. Phys.}\ }\textbf {\bibinfo {volume} {89}},\
  \bibinfo {pages} {015001} (\bibinfo {year} {2017})}\BibitemShut {NoStop}%
\bibitem [{\citenamefont {Weiss}(2012)}]{weiss_quantum_2012}%
  \BibitemOpen
  \bibfield  {author} {\bibinfo {author} {\bibfnamefont {U.}~\bibnamefont
  {Weiss}},\ }\href@noop {} {\emph {\bibinfo {title} {Quantum {{Dissipative
  Systems}}}}}\ (\bibinfo  {publisher} {{World Scientific}},\ \bibinfo {year}
  {2012})\BibitemShut {NoStop}%
\bibitem [{\citenamefont {Albert}\ and\ \citenamefont
  {Jiang}(2014)}]{albert_symmetries_2014-2}%
  \BibitemOpen
  \bibfield  {author} {\bibinfo {author} {\bibfnamefont {V.~V.}\ \bibnamefont
  {Albert}}\ and\ \bibinfo {author} {\bibfnamefont {L.}~\bibnamefont {Jiang}},\
  }\href {\doibase 10.1103/PhysRevA.89.022118} {\bibfield  {journal} {\bibinfo
  {journal} {Phys. Rev. A}\ }\textbf {\bibinfo {volume} {89}},\ \bibinfo
  {pages} {022118} (\bibinfo {year} {2014})}\BibitemShut {NoStop}%
\bibitem [{Note3()}]{Note3}%
  \BibitemOpen
  \bibinfo {note} {The regions are taken here to be disjoint. When the regions
  overlap, fewer points are needed and the numerical effort is
  reduced.}\BibitemShut {Stop}%
\bibitem [{\citenamefont {Baker}, \citenamefont {Jessup},\ and\ \citenamefont
  {Manteuffel}(2005)}]{baker_technique_2005}%
  \BibitemOpen
  \bibfield  {author} {\bibinfo {author} {\bibfnamefont {A.}~\bibnamefont
  {Baker}}, \bibinfo {author} {\bibfnamefont {E.}~\bibnamefont {Jessup}}, \
  and\ \bibinfo {author} {\bibfnamefont {T.}~\bibnamefont {Manteuffel}},\
  }\href {\doibase 10.1137/S0895479803422014} {\bibfield  {journal} {\bibinfo
  {journal} {SIAM J. Matrix Anal. A.}\ }\textbf {\bibinfo {volume} {26}},\
  \bibinfo {pages} {962} (\bibinfo {year} {2005})}\BibitemShut {NoStop}%
\bibitem [{\citenamefont {Pollard}\ and\ \citenamefont
  {Friesner}(1994)}]{pollard_solution_1994}%
  \BibitemOpen
  \bibfield  {author} {\bibinfo {author} {\bibfnamefont {W.~T.}\ \bibnamefont
  {Pollard}}\ and\ \bibinfo {author} {\bibfnamefont {R.~A.}\ \bibnamefont
  {Friesner}},\ }\href {\doibase 10.1063/1.467222} {\bibfield  {journal}
  {\bibinfo  {journal} {J. Chem. Phys}\ }\textbf {\bibinfo {volume} {100}},\
  \bibinfo {pages} {5054} (\bibinfo {year} {1994})}\BibitemShut {NoStop}%
\bibitem [{\citenamefont {Anderson}\ \emph {et~al.}(1999)\citenamefont
  {Anderson}, \citenamefont {Bai}, \citenamefont {Bischof}, \citenamefont
  {Blackford}, \citenamefont {Demmel}, \citenamefont {Dongarra}, \citenamefont
  {Croz}, \citenamefont {Greenbaum}, \citenamefont {Hammerling}, \citenamefont
  {McKenney},\ and\ \citenamefont {Sorensen}}]{anderson_lapack_1999}%
  \BibitemOpen
  \bibfield  {author} {\bibinfo {author} {\bibfnamefont {E.}~\bibnamefont
  {Anderson}}, \bibinfo {author} {\bibfnamefont {Z.}~\bibnamefont {Bai}},
  \bibinfo {author} {\bibfnamefont {C.}~\bibnamefont {Bischof}}, \bibinfo
  {author} {\bibfnamefont {S.}~\bibnamefont {Blackford}}, \bibinfo {author}
  {\bibfnamefont {J.}~\bibnamefont {Demmel}}, \bibinfo {author} {\bibfnamefont
  {J.}~\bibnamefont {Dongarra}}, \bibinfo {author} {\bibfnamefont {J.~D.}\
  \bibnamefont {Croz}}, \bibinfo {author} {\bibfnamefont {A.}~\bibnamefont
  {Greenbaum}}, \bibinfo {author} {\bibfnamefont {S.}~\bibnamefont
  {Hammerling}}, \bibinfo {author} {\bibfnamefont {A.}~\bibnamefont
  {McKenney}}, \ and\ \bibinfo {author} {\bibfnamefont {D.}~\bibnamefont
  {Sorensen}},\ }\href@noop {} {\emph {\bibinfo {title} {{{LAPACK Users}}'
  {{Guide}}, {{Third Edition}}}}}\ (\bibinfo  {publisher} {{SIAM}},\ \bibinfo
  {address} {{Philadelphia, Penn.}},\ \bibinfo {year} {1999})\BibitemShut
  {NoStop}%
\bibitem [{\citenamefont {Paul}\ \emph {et~al.}(2017)\citenamefont {Paul},
  \citenamefont {Sengupta}, \citenamefont {Ark}, \citenamefont {Tu},
  \citenamefont {Zhao},\ and\ \citenamefont {Boppart}}]{paul_coherent_2017}%
  \BibitemOpen
  \bibfield  {author} {\bibinfo {author} {\bibfnamefont {K.}~\bibnamefont
  {Paul}}, \bibinfo {author} {\bibfnamefont {P.}~\bibnamefont {Sengupta}},
  \bibinfo {author} {\bibfnamefont {E.~D.}\ \bibnamefont {Ark}}, \bibinfo
  {author} {\bibfnamefont {H.}~\bibnamefont {Tu}}, \bibinfo {author}
  {\bibfnamefont {Y.}~\bibnamefont {Zhao}}, \ and\ \bibinfo {author}
  {\bibfnamefont {S.~A.}\ \bibnamefont {Boppart}},\ }\href {\doibase
  10.1038/nphys4257} {\bibfield  {journal} {\bibinfo  {journal} {Nat. Phys.}\
  }\textbf {\bibinfo {volume} {13}},\ \bibinfo {pages} {1111} (\bibinfo {year}
  {2017})}\BibitemShut {NoStop}%
\bibitem [{\citenamefont {Raab}\ \emph {et~al.}(1999)\citenamefont {Raab},
  \citenamefont {Worth}, \citenamefont {Meyer},\ and\ \citenamefont
  {Cederbaum}}]{raab_molecular_1999}%
  \BibitemOpen
  \bibfield  {author} {\bibinfo {author} {\bibfnamefont {A.}~\bibnamefont
  {Raab}}, \bibinfo {author} {\bibfnamefont {G.~A.}\ \bibnamefont {Worth}},
  \bibinfo {author} {\bibfnamefont {H.-D.}\ \bibnamefont {Meyer}}, \ and\
  \bibinfo {author} {\bibfnamefont {L.~S.}\ \bibnamefont {Cederbaum}},\ }\href
  {\doibase doi:10.1063/1.478061} {\bibfield  {journal} {\bibinfo  {journal}
  {J. Chem. Phys.}\ }\textbf {\bibinfo {volume} {110}},\ \bibinfo {pages} {936}
  (\bibinfo {year} {1999})}\BibitemShut {NoStop}%
\bibitem [{\citenamefont {Sukharev}\ and\ \citenamefont
  {Seideman}(2004)}]{sukharev_optimal_2004}%
  \BibitemOpen
  \bibfield  {author} {\bibinfo {author} {\bibfnamefont {M.}~\bibnamefont
  {Sukharev}}\ and\ \bibinfo {author} {\bibfnamefont {T.}~\bibnamefont
  {Seideman}},\ }\href {\doibase 10.1103/PhysRevLett.93.093004} {\bibfield
  {journal} {\bibinfo  {journal} {Phys. Rev. Lett.}\ }\textbf {\bibinfo
  {volume} {93}},\ \bibinfo {pages} {093004} (\bibinfo {year}
  {2004})}\BibitemShut {NoStop}%
\bibitem [{\citenamefont {Christopher}, \citenamefont {Shapiro},\ and\
  \citenamefont {Brumer}(2005)}]{christopher_overlapping_2005}%
  \BibitemOpen
  \bibfield  {author} {\bibinfo {author} {\bibfnamefont {P.~S.}\ \bibnamefont
  {Christopher}}, \bibinfo {author} {\bibfnamefont {M.}~\bibnamefont
  {Shapiro}}, \ and\ \bibinfo {author} {\bibfnamefont {P.}~\bibnamefont
  {Brumer}},\ }\href {\doibase 10.1063/1.2000260} {\bibfield  {journal}
  {\bibinfo  {journal} {J. Chem. Phys.}\ }\textbf {\bibinfo {volume} {123}},\
  \bibinfo {pages} {064313} (\bibinfo {year} {2005})}\BibitemShut {NoStop}%
\bibitem [{\citenamefont {Christopher}, \citenamefont {Shapiro},\ and\
  \citenamefont {Brumer}(2006)}]{christopher_quantum_2006}%
  \BibitemOpen
  \bibfield  {author} {\bibinfo {author} {\bibfnamefont {P.~S.}\ \bibnamefont
  {Christopher}}, \bibinfo {author} {\bibfnamefont {M.}~\bibnamefont
  {Shapiro}}, \ and\ \bibinfo {author} {\bibfnamefont {P.}~\bibnamefont
  {Brumer}},\ }\href {\doibase 10.1063/1.2346684} {\bibfield  {journal}
  {\bibinfo  {journal} {J. Chem. Phys.}\ }\textbf {\bibinfo {volume} {125}},\
  \bibinfo {pages} {124310} (\bibinfo {year} {2006})}\BibitemShut {NoStop}%
\bibitem [{\citenamefont {Ryabinkin}, \citenamefont {{Joubert-Doriol}},\ and\
  \citenamefont {Izmaylov}(2014)}]{ryabinkin_when_2014}%
  \BibitemOpen
  \bibfield  {author} {\bibinfo {author} {\bibfnamefont {I.~G.}\ \bibnamefont
  {Ryabinkin}}, \bibinfo {author} {\bibfnamefont {L.}~\bibnamefont
  {{Joubert-Doriol}}}, \ and\ \bibinfo {author} {\bibfnamefont {A.~F.}\
  \bibnamefont {Izmaylov}},\ }\href {\doibase 10.1063/1.4881147} {\bibfield
  {journal} {\bibinfo  {journal} {J. Chem. Phys.}\ }\textbf {\bibinfo {volume}
  {140}},\ \bibinfo {pages} {214116} (\bibinfo {year} {2014})}\BibitemShut
  {NoStop}%
\bibitem [{Note4()}]{Note4}%
  \BibitemOpen
  \bibinfo {note} {Due to the strict time-ordering of the perturbative
  expansion, $I_\protect \text {het,pp}(\omega ; \tau )$ is zero when the probe
  arrives before the pump, i.e. it is background-free.}\BibitemShut {Stop}%
\bibitem [{\citenamefont {Novak}\ and\ \citenamefont
  {Ritter}(1997)}]{novak_curse_1997}%
  \BibitemOpen
  \bibfield  {author} {\bibinfo {author} {\bibfnamefont {E.}~\bibnamefont
  {Novak}}\ and\ \bibinfo {author} {\bibfnamefont {K.}~\bibnamefont {Ritter}},\
  }in\ \href@noop {} {\emph {\bibinfo {booktitle} {Multivariate
  {{Approximation}} and {{Splines}}}}},\ \bibinfo {series and number} {{{ISNM
  International Series}} of {{Numerical Mathematics}}},\ \bibinfo {editor}
  {edited by\ \bibinfo {editor} {\bibfnamefont {G.}~\bibnamefont
  {N{\"u}rnberger}}, \bibinfo {editor} {\bibfnamefont {J.~W.}\ \bibnamefont
  {Schmidt}}, \ and\ \bibinfo {editor} {\bibfnamefont {G.}~\bibnamefont
  {Walz}}}\ (\bibinfo  {publisher} {{Birkh{\"a}user Basel}},\ \bibinfo {year}
  {1997})\ pp.\ \bibinfo {pages} {177--187}\BibitemShut {NoStop}%
\bibitem [{\citenamefont {Torii}(2006)}]{torii_effects_2006}%
  \BibitemOpen
  \bibfield  {author} {\bibinfo {author} {\bibfnamefont {H.}~\bibnamefont
  {Torii}},\ }\href {\doibase 10.1021/jp060014c} {\bibfield  {journal}
  {\bibinfo  {journal} {J. Phys. Chem. A}\ }\textbf {\bibinfo {volume} {110}},\
  \bibinfo {pages} {4822} (\bibinfo {year} {2006})}\BibitemShut {NoStop}%
\bibitem [{\citenamefont {Thirumalai}, \citenamefont {Bruskin},\ and\
  \citenamefont {Berne}(1983)}]{thirumalai_iterative_1983}%
  \BibitemOpen
  \bibfield  {author} {\bibinfo {author} {\bibfnamefont {D.}~\bibnamefont
  {Thirumalai}}, \bibinfo {author} {\bibfnamefont {E.~J.}\ \bibnamefont
  {Bruskin}}, \ and\ \bibinfo {author} {\bibfnamefont {B.~J.}\ \bibnamefont
  {Berne}},\ }\href {\doibase 10.1063/1.445601} {\bibfield  {journal} {\bibinfo
   {journal} {J. Chem. Phys.}\ }\textbf {\bibinfo {volume} {79}},\ \bibinfo
  {pages} {5063} (\bibinfo {year} {1983})}\BibitemShut {NoStop}%
\bibitem [{\citenamefont {Shao}\ and\ \citenamefont
  {Makri}(2001)}]{shao_iterative_2001}%
  \BibitemOpen
  \bibfield  {author} {\bibinfo {author} {\bibfnamefont {J.}~\bibnamefont
  {Shao}}\ and\ \bibinfo {author} {\bibfnamefont {N.}~\bibnamefont {Makri}},\
  }\href {\doibase 10.1016/S0301-0104(01)00286-5} {\bibfield  {journal}
  {\bibinfo  {journal} {Chemical Physics}\ }\textbf {\bibinfo {volume} {268}},\
  \bibinfo {pages} {1} (\bibinfo {year} {2001})}\BibitemShut {NoStop}%
\bibitem [{\citenamefont {Jadhao}\ and\ \citenamefont
  {Makri}(2008)}]{jadhao_iterative_2008}%
  \BibitemOpen
  \bibfield  {author} {\bibinfo {author} {\bibfnamefont {V.}~\bibnamefont
  {Jadhao}}\ and\ \bibinfo {author} {\bibfnamefont {N.}~\bibnamefont {Makri}},\
  }\href {\doibase 10.1063/1.3000393} {\bibfield  {journal} {\bibinfo
  {journal} {J. Chem. Phys.}\ }\textbf {\bibinfo {volume} {129}},\ \bibinfo
  {pages} {161102} (\bibinfo {year} {2008})}\BibitemShut {NoStop}%
\bibitem [{\citenamefont {Wiener}(1930)}]{wiener_generalized_1930}%
  \BibitemOpen
  \bibfield  {author} {\bibinfo {author} {\bibfnamefont {N.}~\bibnamefont
  {Wiener}},\ }\href {\doibase 10.1007/BF02546511} {\bibfield  {journal}
  {\bibinfo  {journal} {Acta Math.}\ }\textbf {\bibinfo {volume} {55}},\
  \bibinfo {pages} {117} (\bibinfo {year} {1930})}\BibitemShut {NoStop}%
\bibitem [{\citenamefont {Rudin}(1987)}]{rudin_real_1987}%
  \BibitemOpen
  \bibfield  {author} {\bibinfo {author} {\bibfnamefont {W.}~\bibnamefont
  {Rudin}},\ }\href@noop {} {\emph {\bibinfo {title} {Real and Complex
  Analysis}}}\ (\bibinfo  {publisher} {{McGraw-Hill}},\ \bibinfo {year}
  {1987})\BibitemShut {NoStop}%
\end{thebibliography}%
